\documentclass[twocolumn,showpacs,preprintnumbers,amsmath,amssymb,floatfix,prd]{revtex4}
\usepackage{epsfig}

\begin{document}
\title{Fragmentation Functions in Nuclear Media}
\author{Rodolfo Sassot}\email{sassot@df.uba.ar}
\affiliation{Instituto de F\'{\i}sica de Buenos Aires, CONICET, \\ Departamento de F\'{\i}sica, 
Facultad de Ciencias Exactas y Naturales, Universidad de Buenos Aires, Ciudad Universitaria, Pabell\'on\ 1 (1428)
Buenos Aires, Argentina}
\author{Marco Stratmann}\email{marco@ribf.riken.jp}
\affiliation{Institut f\"{u}r Theoretische Physik,
Universit\"{a}t Regensburg, 93040 Regensburg, Germany \\
Institut f\"{u}r Theoretische Physik und Astrophysik, Universit\"{a}t W\"{u}rzburg,
97074 W\"{u}rzburg, Germany}
\author{Pia Zurita}\email{pia@df.uba.ar}
\affiliation{Instituto de F\'{\i}sica de Buenos Aires, CONICET, \\ Departamento de F\'{\i}sica, 
Facultad de Ciencias Exactas y Naturales, Universidad de Buenos Aires, Ciudad Universitaria, Pabellon\ 1 (1428)
Buenos Aires, Argentina}

\begin{abstract}
We perform a detailed phenomenological analysis of how well hadronization in nuclear environments 
can be described in terms of effective fragmentation functions.  
The medium modified fragmentation functions are assumed to factorize from the 
partonic scattering cross sections and evolve in the
hard scale in the same way as the standard or vacuum fragmentation functions.
Based on precise data on semi-inclusive deep-inelastic scattering off nuclei
and hadron production in deuteron-gold collisions, we extract 
sets of effective fragmentation functions for pions and kaons at NLO accuracy. 
The obtained sets provide a rather accurate description of the kinematical dependence 
of the analyzed cross sections and are found to differ 
significantly from standard fragmentation functions both in shape and magnitude. 
Our results support the notion of factorization and universality in the studied nuclear
environments, at least in an effective way and within the precision of the available data. 
\end{abstract}


\maketitle

%%%%%%%%%%%%%%%%%%%%%%%%%%%%%%%%%%%%%
\section{Introduction and Motivation}
%%%%%%%%%%%%%%%%%%%%%%%%%%%%%%%%%%%%%
In spite of the remarkable phenomenological success of Quantum
Chromodynamics (QCD) as the theory of strong 
interactions, a detailed quantitative understanding of hadronization processes 
is still one of the great challenges for the theory.
Hadronization is the mechanism by which a final-state quark or gluon, 
excited in some hard partonic interaction, dresses itself and develops into an 
observed final-state hadron. As such, it is sensitive to physics happening
at long distances and time scales, and perturbative tools that have earned 
QCD its present standing, simply fall short. On the other hand, studying
hadronization process is a path to understand the phenomenon of confinement 
and opens a window to the non-perturbative domain of QCD.

The last few years have seen a significant improvement in the 
perturbative QCD (pQCD) description of scattering processes 
with identified produced hadrons.
More specifically, precise determinations of fragmentation functions
for various different hadron species have been performed 
\cite{deFlorian:2007,Hirai:2007cx,Albino:2008fy}.
Fragmentation functions parametrize the non-perturbative details of the 
hadronization process and, by virtue of the factorization theorem, are
assumed to be universal, i.e., process independent \cite{ref:framework}.
One of the most interesting results of these studies is that the standard pQCD framework not 
only reproduces the data on electron-positron annihilation into hadrons 
with remarkable precision, but describes equally well other processes, 
like semi-inclusive deep-inelastic scattering (SIDIS) and hadron production 
in proton-proton collisions \cite{deFlorian:2007}.
Using the wealth of data with identified hadrons in a global QCD analysis
not only increases the precision and kinematical coverage
of fragmentation functions, but more importantly, 
supports the ideas of factorization and universality 
in the kinematical domain accessed by these experiments.
These concepts are the starting points for the successful 
pQCD description of hard scattering processes.

For more than thirty years, it is known that results for hadron production processes
occurring in a nuclear environment can differ significantly from similar experiments
involving only light nuclei or proton targets \cite{ref:oldexp}.
More recently, detailed kinematic distributions for production rates 
of identified hadrons in SIDIS off different nuclei
have been provided by the HERMES experiment
\cite{Airapetian:2007vu}.
The origin of the observed differences induced by the nuclear media,
has been attributed to a variety of 
conceivable mechanisms besides the well-known modification of parton densities in nuclei.
Ideas range from interactions between the nuclear medium 
and the seed partons before the hadronization takes place, as implemented 
through pQCD inspired models, to interactions between the medium and the 
final-state hadrons, usually formulated in a nuclear and hadron physics language;
for recent reviews, see Refs.~\cite{Arleo:2008dn,Accardi:2009qv}. 
Most models reproduce, with different degree of success, some features of the data, in 
spite of very different, even orthogonal theoretical approaches and ingredients.

In the case of the so called initial-state nuclear effects, such as those observed in
inclusive deep-inelastic scattering (DIS) off nuclei \cite{ref:ndis} 
and the Drell-Yan process on nuclear targets \cite{ref:ndy}, 
it has been shown that the influence of the nuclear environment 
can be effectively factorized into a set of modified nuclear parton distribution functions (nPDFs).
The nPDFs scale in energy with standard evolution equations, and conventional
partonic hard scattering cross sections can be utilized in calculations.
Within the precision of the available data, this approach has been demonstrated 
to be an excellent approximation at next-to-leading order (NLO) accuracy of pQCD, 
and allows one to continue to exploit the key features of factorization and 
universality in a nuclear environment \cite{deFlorian:2003qf,Hirai:2007sx,Eskola:2009uj}. 
From the QCD point of view, the nPDFs carry all the non-perturbative information
related to the probed nuclei, just as the standard parton densities parametrize
our knowledge of the nucleon. Both need to be obtained in global QCD fits to data 
or approximated within a given non-perturbative model. 

It is quite natural then to ask if the idea of factorization can be extended to 
final-state nuclear effects and to explore how good such an approximation works in practice
or, alternatively, to determine where and why it breaks down.
From a theoretical point of view, however, the answer to the question  
whether the introduction of medium modified fragmentation functions (nFFs)
is a viable concept is not obvious. On the one hand, interactions with the 
nuclear medium may spoil the requirements of factorization theorems, 
but on the other hand any estimate of the factorization breaking effects 
is strongly model dependent. 

Assuming that factorization holds, the medium modified fragmentation functions 
should contain (factorize) all the non-perturbative details relevant for 
the computation of hard processes with identified hadrons and would be 
interchangeable from one process to another, i.e., universal.
On top of that, such a theoretical framework is well defined,
has predictive power, and can be systematically extended beyond the leading
order (LO) approximation.
The inclusion of higher order QCD corrections and the possibility to use different 
observables simultaneously in a global analysis have both been proven
to be crucial for a successful extraction of vacuum fragmentation functions (FFs)
\cite{deFlorian:2007}.
All these aspects can be explicitly tested also in a nuclear environment,
using data from an increasing number of experiments that have performed 
very precise measurements of inclusive hadron production off nuclear targets.
They comprise semi-inclusive measurements of hadron multiplicities
by HERMES \cite{Airapetian:2007vu}, as well as experiments 
in deuteron-gold ($dAu$) collisions at the BNL-RHIC \cite{ref:phenix,ref:star,ref:star-thesis}. 
Both processes show clear signals of a non-trivial nuclear dependence in the hadronization mechanism. 

In addition to our primary goal, which is testing the factorization properties of nFFs
in a consistent theoretical framework,
we also aim to constrain them as precisely as possible 
from the different data sets currently available.
This will allow one to compare our results for the nFFs with the different model estimates
and mechanisms proposed to model hadronization in a nuclear medium
\cite{Arleo:2008dn,Accardi:2009qv}. 
Together with our analysis of vacuum FFs \cite{deFlorian:2007}, the present work also serves as baseline 
for ongoing studies of processes with detected hadrons 
in heavy ion collisions performed at BNL-RHIC and at the CERN-LHC in the future. 
These projects are aimed at investigating the properties of hot and dense QCD matter, 
and require precise knowledge on hadronization under both normal and 
such extreme conditions \cite{Armesto:2009ug}.

The paper is organized as follows:
in the next Section, we very briefly summarize the pQCD framework for FFs and discuss 
how to extend it to nFFs, specifically for the processes on which we will base our analysis. 
We also review the different aspects of the data that suggest medium induced effects
in fragmentation processes. 
In Sec.~III, we first show how the main features of the experimental results
can be reproduced in a very basic scheme of medium modifications, 
associated with a simple parameterization of nFFs which can be motivated by intuitive arguments.
Next, we present our results of a more refined determination of the
medium modified fragmentation functions for pions at NLO accuracy. 
We briefly comment on different centrality classes in deuteron-gold collisions at RHIC. 
Finally, we extend our analysis to measurements of kaon production in a nuclear environment. 
We summarize our results in Sec.~IV.

%%%%%%%%%%%%%%%%%%%%%%%%%%%%%%%%%%%%%%%%%%%%%%%%%%%%%%%%%%%%%%
\section{\label{sec:framework}QCD Framework for 
medium modified  fragmentation functions}
%%%%%%%%%%%%%%%%%%%%%%%%%%%%%%%%%%%%%%%%%%%%%%%%%%%%%%%%%%%%%%

%%%%%%%%%%%%%%%%%%%%%%%%%%%%%%%%%%%%%%%%%%%%%%%%%%
\subsection{\label{sec:basic} Basic properties of FFs}
%%%%%%%%%%%%%%%%%%%%%%%%%%%%%%%%%%%%%%%%%%%%%%%%%%
%
In the naive parton model, fragmentation functions $D_{i}^{H}(z)$ are simply taken 
as the probabilities for a final-state parton of a given flavor $i$ 
to produce a specific hadron $H$, carrying a fraction $z$ of its four-momentum. 
This intuitive picture can be consistently extended to the field-theoretical
language of QCD by defining fragmentation functions in terms of bilocal operators 
in a certain factorization scheme \cite{ref:collins-soper}. 
The most common choice for the latter is the $\overline{\mathrm{MS}}$ 
scheme, which we also adopt throughout this work.
Along with analogous operator definitions for the parton distribution functions (PDFs),
the factorization of short- and long-distance contributions to one-hadron 
inclusive cross sections is then precisely determined, and higher order QCD corrections
can the be systematically taken into account. 
While the non-perturbative but universal FFs (and PDFs) need to be extracted from 
data through global QCD analyses, both the relevant hard partonic scattering 
cross sections and the scale evolution of FFs (and PDFs) are calculable within pQCD.
Corrections to this factorized framework emerge as inverse powers of the
hard scale characterizing the one-hadron inclusive cross section, 
like the virtuality $Q^2$ of the exchanged virtual
photon in $e^+e^-$ annihilations or the transverse momentum $p_T$ of the
produced hadron in proton-proton collisions. Provided that $Q^2$ or $p_T$
are large enough, these higher twist contributions can be safely neglected in
phenomenological applications.

The scale dependence of the $D_{i}^{H}$ is governed by a set of
coupled renormalization group equations very similar to those
for PDFs. They schematically read \cite{ref:rijken,ref:nlo-kernels}:
\begin{eqnarray}
\label{eq:singevol}
\frac{d}{d\ln Q^2} \vec{D}^H (z,Q^2) = \left[
\hat{P}^{(T)}\otimes \vec{D}^H\right](z,Q^2),
\end{eqnarray}
where
\begin{equation} \label{eq:singlet}
\vec{D}^H \equiv \left( \begin{array}{c} D_{\Sigma}^H \\ D_g^H \\
\end{array}\right),\,\,\,
D_{\Sigma}^H \equiv \sum_q (D_q^H+ D_{\bar{q}}^H)
\end{equation}
and
\begin{eqnarray}
\label{eq:pmatrix}
\renewcommand{\arraystretch}{1.3}
\hat{P}^{(T)} \equiv \left( \begin{array}{cc}
P_{qq}^{(T)} &  2n_f P_{gq}^{(T)} \\
\frac{1}{2n_f} P_{qg}^{(T)} & P_{gg}^{(T)} \\
\end{array}\right) 
\end{eqnarray}
is the matrix of the singlet {\em timelike} evolution kernels.
The symbol $\otimes$ denotes a standard convolution.
The NLO splitting functions $P_{ij}^{(T)}$ have been computed in
\cite{ref:nlo-kernels,ref:rijken} or can be related to the
corresponding spacelike kernels by proper
analytic continuation \cite{ref:sv-kernels}.
Recently, the diagonal splitting functions
$P_{qq}^{(T)}$ and $P_{gg}^{(T)}$ have been calculated
up to next-to-next-to-leading order accuracy \cite{ref:nnlo}.

Apart from the requirement of a sufficiently large hard scale
to suppress power corrections, the applicability of FFs is in practice
further limited to the range of medium-to-large momentum fractions $z$
\cite{deFlorian:2007,Hirai:2007cx,Albino:2008fy}.
While large perturbative contributions to the 
splitting kernels $P_{gq}^{(T)}$ and $P_{gg}^{(T)}$ 
at small $z$ can be resummed to all orders, there is no systematic 
or unique way to include corrections related to the ignored mass of
the produced hadron $H$. In any case, ``resummed'' or ``mass corrected''
FFs obtained in one process should not be used with fixed order
expressions in other processes. However, it is a key point of
global QCD analyses of FFs to combine diverse sets of data 
in their determination \cite{deFlorian:2007}.
Only by consistently exploiting 
a large variety of one-hadron inclusive processes one can
test the underlying assumption of factorization, implying  
that FFs are exactly the same no matter if the seed parton 
is excited from the vacuum like in $e^+e^-$ annihilation 
or from a nucleon in the case of one-particle-inclusive 
processes in either lepton-nucleon or nucleon-nucleon collisions.
To avoid problems at small $z$, we follow Ref.~\cite{deFlorian:2007}
and simply impose a cut $z>z_{\min}=0.05 (0.1)$ on all data 
with identified pions (kaons) used in our analysis
of medium modified fragmentation functions.

%%%%%%%%%%%%%%%%%%%%%%%%%%%%%%%%%%%%%%%%%%%%%%%%%%
\subsection{\label{sec:ansatz} Convolution approach for medium modified FFs}
%%%%%%%%%%%%%%%%%%%%%%%%%%%%%%%%%%%%%%%%%%%%%%%%%%
%
Assuming that hard collinear factorization for one-particle inclusive processes
is realized also for collisions involving nuclei, one can define 
fragmentation functions $D_{i/A}^{H}(z)$ specific for nuclear environments
characterized by their atomic number $A$.
Like standard FFs, the medium modified $D_{i/A}^{H}(z)$ describe the hadronization
of a parton $i$ into a hadron $H$ but now in the background of a nucleus $A$.
Due to medium-induced final-state soft exchanges which happen after the hard
partonic scattering process, the non-perturbative content of the $D_{i/A}^{H}(z)$
could differ significantly from the one known for vacuum fragmentation functions
$D_{i}^{H}(z)$. A similar reasoning is applied with great phenomenological success
in analyses of nPDFs that account for medium-induced effects 
in the initial-state \cite{deFlorian:2003qf,Hirai:2007sx,Eskola:2009uj}.

Provided that both FFs and nFFs have the same factorization properties, 
we can use exactly the same theoretical expressions in computations of measured
hadron yields for processes involving free or nuclear bounded nucleons,
replacing only the PDFs and FFs by the corresponding nPDFs and nFFs in the latter case.
The global analysis of the nFFs, that is the precise 
determination of the medium induced modifications to FFs
based on the presently available data,
proceeds in close analogy to those for FFs. 
The stringent framework of pQCD and
factorization does not require to make any specific assumptions on the
size and sign of nuclear modifications prior to the fit, 
and the results are entirely determined by data.
Known nuclear effects on parton densities, as seen, e.g., in
DIS off nuclei \cite{ref:ndis}, are fully accounted for by choosing a
recent set of nPDFs. 
The obtained sets of nFFs will allow us to test how well the 
underlying assumption of factorization works in practice and
can be compared to the different ideas and mechanisms
for medium modifications of quark and gluon FFs proposed in the literature
\cite{Arleo:2008dn,Accardi:2009qv}.
We refer the reader to Ref.~\cite{deFlorian:2007} for detailed expressions of the 
relevant cross sections and an outline of the numerical fitting procedure.

Rather than fitting from scratch the nFFs, which would take as many parameters as the 
standard or vacuum FFs, plus several more to represent their nuclear $A$ dependence,
we choose to relate the $D^{H}_{i/A}$ to the standard ones $D^{H}_{i}$ 
at a given initial scale $Q_0=1\,\mathrm{GeV}$ by a convolution approach:
\begin{equation}
\label{eq:ansatz}
D^{H}_{i/A}(z,Q^2_0) =\int_z^1 \frac{dy}{y}\, W_i^H(y,A,Q^2_0)\, D^{H}_{i}(\frac{z}{y},Q^2_0)\,. 
\end{equation} 
The weight function $W_i^H(y,A,Q^2_0)$ parameterizes all nuclear modifications 
and, at the same time, retains the information already available on the vacuum FFs
for $A=1$. Here we take the NLO sets of DSS \cite{deFlorian:2007} as reference,
which provide an excellent global description of hadron yields in $e^+e^-$, $ep$, and $pp$
processes. We refrain from performing our analysis at leading order accuracy 
which, at best, can give a rough qualitative result. The LO FFs in Ref.~\cite{deFlorian:2007}
yield a significantly less favorable description of the data.
Provided that the functional form for $W_i^H(y,A,Q^2_0)$ is flexible enough,
one can accommodate the specific details of each individual measurement, and the nFFs can
be extracted as precisely as possible from data.
The scale dependence of the nFFs in Eq.~(\ref{eq:ansatz}) is dictated by factorization and
determined by the standard evolution equations for vacuum FFs discussed in Sec.~\ref{sec:basic}.

The impact of the weight function $W_i^H(y,A,Q^2_0)$ in Eq.~({\ref{eq:ansatz})
on the resulting nFFs can be readily understood. A simple Dirac delta function $\delta(1-y)$ as weight would imply no 
medium induced effects in the hadronization process, while a shift in its argument, i.e.,
$\delta(1-\epsilon-y)$, leads to a shift in the momentum fraction $z$ 
as suggested, for instance, by many parton energy loss scenarios \cite{Arleo:2008dn,Accardi:2009qv}. 
A more flexible weight function like
\begin{equation}
\label{eq:w-example}
W_i^H(y,A,Q^2_0)=n_i \,y^{\alpha_i}(1-y)^{\beta_i},
\end{equation}
can be used to parameterize effects not necessarily related to partonic mechanisms, 
such as hadron or pre-hadron attenuation or enhancement, with a great economy of parameters.
As will be shown below, the $A$ dependence of the weight function $W_i^H$ can easily be included 
in its coefficients, e.g., $n_i$, $\alpha_i$, and $\beta_i$ in (\ref{eq:w-example}),
by taking them as smooth functions of a nuclear property like its volume, radius, 
or mean density. 

We note that convolutional integrals like in Eq.~(\ref{eq:ansatz})
are the most natural language for parton dynamics at LO and beyond,
showing up ubiquitously in evolution equations, cross sections, etc.
They can be most straightforwardly handled in Mellin moment space,
which is also well suited for a numerically efficient global QCD analysis
\cite{ref:mellin}.
The convolution approach has been demonstrated to be effective and
phenomenologically successful in the extraction of initial-state nuclear effects 
for PDFs at NLO accuracy in Ref.~\cite{deFlorian:2003qf}. For consistency,
we take these sets of nPDFs, in the following labeled as nDS, in all calculations
of cross sections relevant for our global analysis of nFFs.

%%%%%%%%%%%%%%%%%%%%%%%%%%%%%%%%%%%%%%%%%%%%%%%%%%
\subsection{\label{sec:data} Data sensitive to nFFs}
%%%%%%%%%%%%%%%%%%%%%%%%%%%%%%%%%%%%%%%%%%%%%%%%%%
%
The gross features of FFs are usually determined from 
very precise $e^+e^-$ annihilation data, mainly taken by the CERN-LEP experiments.
The lack of this source of information for medium modified fragmentation
functions is evident and significantly complicates their analysis.
All available probes require a careful deconvolution of nFFs from medium effects
related to nPDFs, which can be done consistently in a global QCD analysis based on
factorization. 

One of the most interesting pieces of evidence on medium induced effects in the hadronization
process comes from semi-inclusive deep-inelastic scattering off nuclear 
targets. Such kind of measurements have been performed since the late seventies by different 
collaborations \cite{ref:oldexp} and in recent years have reached a level of 
precision and sophistication
that allows for very detailed quantitative analyses.
Specifically, the HERMES collaboration has performed a series of measurements on deuterium, 
helium, neon, krypton, and xenon targets, with identified charged and neutral pions, kaons, and 
(anti)protons in the final-state \cite{Airapetian:2007vu}. 
The data are presented as distributions in the relevant kinematical variables, such as the 
hadron momentum fraction $z$ and the photon virtuality $Q^2$, which are 
used to characterize fragmentation functions,
as well as the virtual photon energy $\nu$, that can 
be related to the nucleon momentum fraction $x$ carried by initial-state parton.
In addition, data are available in terms of $p_{T}^2$,
the transverse momentum squared of the observed hadron.
The detailed kinematical dependence of the HERMES data \cite{Airapetian:2007vu}, 
combined with the information for different final-state hadrons and target nuclei, 
puts very sharp constraints on the effective,
medium modified fragmentation functions we wish to determine. 

In order to minimize initial-state medium induced effects, which in a factorized approach 
should be contained in the nPDFs, the data are presented as ratios of hadron multiplicities
for heavy nuclei $A$ and deuterium $(d)$,
\begin{equation}
\label{eq:multiplicity}
R^H_A(\nu,Q^2,z,p_T^2)=\frac{\left(\frac{N^H(\nu,Q^2,z,p_T^2)}{N^e(\nu,Q^2)}\right)_A}
{\left(\frac{N^H(\nu,Q^2,z,p_T^2)}{N^e(\nu,Q^2)}\right)_d}\;.
\end{equation}
$N^H(\nu,Q^2,z,p_T^2)$ denotes the number of hadrons of type $H$ produced in SIDIS, 
and $N^e(\nu,Q^2)$ is the number of inclusive leptons in DIS. 
The cancellation of initial-state nuclear effects would be exact in a LO framework 
if medium induced modifications to the PDFs could be represented by a single multiplicative factor 
irrespective of the parton flavor. 
Even though the different parton species are known to require different, non-trivial modifications
\cite{deFlorian:2003qf}, these differences get diluted even at NLO accuracy. 
As will be shown below, they cancel in the ratio (\ref{eq:multiplicity}) to a very good approximation. 

The second crucial piece of evidence on medium induced effects in the hadronization process
comes from single-inclusive identified hadron yields obtained in $dAu$ collisions at mid-rapidity
by the RHIC experiments at BNL \cite{ref:phenix,ref:star,ref:star-thesis}.
These measurements are often seen as ``control experiments" associated with the program of
colliding two heavy-ion beams at RHIC to explore the properties of nuclear matter under extreme conditions. 
However, in the face of the clear evidence of strong medium induced effects 
in the fragmentation process deduced from the SIDIS data \cite{Airapetian:2007vu}, 
the $dAu$ data \cite{ref:phenix,ref:star,ref:star-thesis} also acquire a particular relevance for our analysis.

Nuclear effects on hadron production in $dAu$ collisions are often quantified through
comparison to the corresponding $pp$ spectrum, 
scaled by the average nuclear overlap or geometry function ${\cal{N}}$.
The so defined nuclear modification factor $R_{dAu}^H$ 
\cite{ref:phenix,ref:star,ref:star-thesis} reflects not only the 
medium induced effects on the fragmentation process, 
but also on the PDFs and includes possible deviations due to isospin considerations.
${\cal{N}}$ counts the number of underlying inelastic nucleon-nucleon (binary) collisions
and depends on Glauber model calculations \cite{ref:phenix,ref:star,ref:star-thesis}. Experimental results 
for nucleon modification factors and invariant cross sections are divided into different
centrality classes or presented for the combined minimum bias sample.

Surprisingly, instead of the pronounced hadron attenuation seen in nuclear SIDIS data
\cite{Airapetian:2007vu}, i.e., $R^H_A<1$, the nuclear modification factors in $dAu$ collisions 
show a curious pattern enhancement and suppression depending on $p_T$ \cite{ref:phenix,ref:star,ref:star-thesis}.
The interpretation of these results is complicated by the large amount of
contributing partonic subprocesses with either quark or gluon fragmentation into
the observed hadron $H$. Their individual share to the hadron yield is 
strongly correlated with $p_T$ as will be shown in some detail below.

Rather than using the invariant cross sections, whose theoretical estimates 
suffer from a sizable dependence on the choice of the factorization scale $\mu_f$
\cite{ref:pionnlo}, 
we perform the global fit in terms of the ratio of the $dAu$ minimum bias cross section 
to the corresponding invariant hadron yield in $pp$ collisions,
\begin{equation}
\label{eq:rhicratio}
R^H_{\sigma}(A,p_T)\equiv \frac{1}{2\,A}
\frac{\left. E\,d^3\sigma^H/dp^3\right|_{dA}}
{\left. E\,d^3\sigma^h/dp^3\right|_{pp}}\,,
\end{equation}
normalized by the number of participating nucleons.
The advantage of $R^H_{\sigma}$ is that both numerator and denominator can be computed 
consistently within pQCD at NLO accuracy using 
current sets of PDFs \cite{ref:unpolpdf}, nPDFs \cite{deFlorian:2003qf}, 
and FFs \cite{deFlorian:2007}. 
In addition, the dependence on $\mu_f$ as well as normalization errors associated with 
inaccuracies in determinations of the PDFs and FFs 
tend to cancel in the ratio (\ref{eq:rhicratio}).
As compared to the related nuclear modification factor $R_{dAu}^H$, 
the ratio $R^H_{\sigma}$ is less sensitive to estimates based on
model dependent calculations.
It is worthwhile noticing that the DSS set of FFs \cite{deFlorian:2007} 
was obtained from a global analysis using the same $pp$ collision data 
we use to define the ratios in Eq.~(\ref{eq:rhicratio}). 

The convoluted way in which the information on quark and gluon fragmentation is encoded
in $dAu$ hadron production data is not an obstacle in a factorized approach 
since we can explicitly compute the relevant cross sections in terms of nPDFs and nFFs, 
as we do for SIDIS. In our global analysis, both sets of data will be treated
simultaneously at NLO accuracy, and the optimum set of quark and gluon nFFs will be extracted. 
In this way we can test whether the hadron attenuation found in SIDIS off nuclei can
be matched with the complicated pattern of enhancement and suppression observed in $dAu$
collisions.

The fact that SIDIS is strongly dominated by quark fragmentation suggests that 
quark nFFs will be suppressed. Likewise, from the known dominance of gluons in $pp$
collisions at RHIC in the relevant low-to-medium $p_T$ region \cite{ref:pionnlo}, one expects that
the main medium induced effect for them would be an enhancement.
In the following we describe the details of our global analysis of
nFFs, present the obtained $z$ and $A$ dependence of quark and gluon nFFs
for pions and kaons, and show that the above, naive expectations are qualitatively correct.

The obtained nFFs apply to the production of hadrons in processes 
where a nucleus collides with a lepton, a nucleon, or a very light nucleus like deuterium.
For collisions between two heavy nuclei the distinctive properties of the
created hot and dense matter presumably may emphasize very different effects in the 
hadronization which perhaps even break factorization. This is beyond the scope 
of this first analysis of nFFs and requires more detailed studies in the future.

%%%%%%%%%%%%%%%%%%%%%%%%%%%%%%%%%%%%%%%%%%%%%%%%%%
\section{\label{sec:results} Results}
%%%%%%%%%%%%%%%%%%%%%%%%%%%%%%%%%%%%%%%%%%%%%%%%%%
%
%%%%%%%%%%%%%%%%%%%%%%%%%%%%%%%%%%%%%%%%%%%%%%%%%%%%%%%%%%%%%%%%%%%%
\subsection{\label{sec:weight} Determination of the weight functions}
%%%%%%%%%%%%%%%%%%%%%%%%%%%%%%%%%%%%%%%%%%%%%%%%%%%%%%%%%%%%%%%%%%%%
%
In order to give a first impression of the main features 
of medium induced effects in the hadronization process suggested 
by the SIDIS \cite{Airapetian:2007vu} and $dAu$ \cite{ref:phenix,ref:star,ref:star-thesis} data, 
we start with an extremely simple
assumption for the functional form of the weight
functions $W_i^H(y,A,Q_0^2)$ in Eq.~(\ref{eq:ansatz}).
This will also illustrate the use and convenience of the convolution approach
outlined in Sec.~\ref{sec:ansatz}.

As we have mentioned in Sec.~\ref{sec:data}, the HERMES 
SIDIS data show a $z$ dependent hadron attenuation $R^H<1$ that increases with $A$
\cite{Airapetian:2007vu}. 
Since the $R^H$ do not significantly depend on the different pion charges,
$H=\{\pi^-, \pi^0, \pi^+\}$, it is natural to assume
the same medium effect on the fragmentation probability for both quarks 
and antiquarks. The observed reduction is most economically
implemented by a weight function
\begin{equation}
W_q^{\pi}(y,A,Q^2_0)=n_q \delta(1-y) + \epsilon_q\, \delta(1-\epsilon_q-y)
\label{eq:toy}
\end{equation} 
that depends only on two free parameters $n_q$ and $\epsilon_q$.
The changes induced in the quark fragmentation functions by such a weight 
are readily interpreted as the superposition of two mechanisms.
The first term in (\ref{eq:toy}) leads to a $z$ independent overall reduction
of the nFFs relative to the vacuum FFs, with $n_q$ decreasing from unity with 
nuclear size $A$. 
The second term in (\ref{eq:toy}) takes care of any non-trivial $z$
dependence in $R^H$ in the most simple way by shifting the
effective value of $z$ probed in the nuclear medium by a small amount $\epsilon_q$
which grows from zero with $A$. Such a shift is also suggested by models
based on parton energy {\em loss}.

For gluons, one could try a similar weight function, however, low $p_T$ pion yields 
in $dAu$ collisions \cite{ref:phenix,ref:star,ref:star-thesis}, 
where gluons are known to dominate in $pp$ processes \cite{ref:pionnlo}, 
suggest that the main effect should be an effective enhancement 
of the fragmentation probability rather than a reduction.
Therefore $n_g$ is most likely to be different from $n_q$.
Other than in SIDIS, $dAu$ data provide only information on the
$z$ dependence in a convoluted way, i.e., integrated over a certain range in $z$.
Because of this, and to a first approximation, we simply take $\epsilon_g=\epsilon_q$.

%
%%%%%%%%%%%%%%
% TABLE 1
%%%%%%%%%%%%%%
%
\begin{table}[th]
\caption{\label{tab:toypara}Coefficients parametrizing the 
naive weight functions $W_{q,g}^{\pi}$ in Eq.~(\ref{eq:toy})
at the input scale $Q_0=1\,\mathrm{GeV}$ for different nuclei $A$.
The three fitted parameters in Eq.~(\ref{eq:toy-adep}) 
are shown in the last line.}
\begin{ruledtabular}
\begin{tabular}{cccc}
 A  & $n_q$ & $\epsilon_q=\epsilon_g$ & $n_g$   \\ \hline
 He & 0.966 & 0.001      & 1.015 \\
 Ne & 0.902 & 0.002      & 1.044 \\  
 Kr & 0.745 & 0.006      & 1.115 \\ 
 Xe & 0.657 & 0.008      & 1.155 \\  
 Au & 0.550 & 0.010      & 1.203 \\  \hline
   & $\gamma_{n_q}=-0.0133$ & $\gamma_{\epsilon}=0.003$ &$\gamma_{n_g}=0.006$
\end{tabular}
\end{ruledtabular}
\end{table}
The nuclear $A$ dependence of the coefficients $n_{q,g}$ and
$\epsilon_{q,g}$ parametrizing the naive weight functions 
$W_{q,g}^{\pi}$ in Eq.~(\ref{eq:toy})
can be easily implemented by a simple ansatz 
\begin{eqnarray}
\label{eq:toy-adep}
n_q&=&1+\gamma_{n_q}\,A^{2/3} \;,\nonumber \\
n_g&=&1+\gamma_{n_g}\,A^{2/3} \;,\\
\epsilon_q&=&\epsilon_g=\gamma_{\epsilon}\,A^{2/3}\;.\nonumber 
\end{eqnarray}
Such an $A$ dependence could be motivated by the way in which the 
volume of a disk with nuclear radius $r_A\simeq r_0\,A^{1/3}$, scales with $A$. 
In total this results in only three free parameters, 
$\{\gamma_{n_q}$, $\gamma_{n_g}$, $\gamma_{\epsilon}\}$,
to be determined by the global fit.

As we will demonstrate in detail below, this very simple minded parameterization 
(in the following labeled as nFF*) reproduces to a good approximation the 
normalization and general trend of pion yields in SIDIS off nuclei and in 
$dAu$ collisions. For completeness, 
the obtained coefficients parameterizing the naive weight
functions are summarized in Tab.~\ref{tab:toypara}.
The ansatz in Eqs.~(\ref{eq:toy}) and (\ref{eq:toy-adep})
is, however, not flexible enough to satisfactorily describe more detailed features of the data, 
for instance, regarding the $x$ dependence of $R^{\pi}_A$ in SIDIS or the
$p_T$ dependence of the $dAu$ data over the entire range.
This is also reflected in the total $\chi^2$ (to be defined below) 
per degree of freedom (d.o.f.) of such a fit, which is close to 2.

These shortcomings suggest that we need to implement much more flexible 
weight functions to fully exploit the constraining power of the data. 
In principle, these weights $W^H_i$ could represent underlying mechanisms 
whose consequences would be other than overall changes in normalization 
and shifts in momentum fraction $z$, but still preserving factorization.
Specifically, we adopt the following ansatz for our global analysis of
nFFs for pions and kaons in Eq.~(\ref{eq:ansatz}),
\begin{eqnarray}
\nonumber
W_q^H(y,A,Q^2_0)&=&n_q \, y^{\alpha_q} (1-y)^{\beta_q} +n'_q\delta(1-\epsilon_q-y)\,,\\
\nonumber
W_g^H(y,A,Q^2_0)&=&n_g \, y^{\alpha_g} (1-y)^{\beta_g} +n'_g\delta(1-\epsilon_g-y)\,,\\
\label{eq:flexi}
\end{eqnarray}
discriminating between quarks and gluons. With the currently available data, 
no significant improvement of the fit is found by introducing an additional 
weight function for antiquarks different from the one used for quarks. 
In addition, the sensitivity of the data on 
the precise $z$ dependence of the medium modified gluon fragmentation
is not sufficient to allow for independent shifts $\epsilon_q$ and $\epsilon_g$
for quarks and gluons, respectively. Like in our simple ansatz in
Eq.~(\ref{eq:toy}), we set $\epsilon_q=\epsilon_g$, but we allow for $n'_q \neq n'_g$. 
The parameterizations of the weight functions in Eq.~(\ref{eq:flexi}) 
can be seen as a natural extension of Eq.~(\ref{eq:toy}), where
the first term can now have a flexible $z$ dependence, and 
free coefficients $n_i'$ add extra flexibility to the second, energy loss term. 

The nuclear dependence of the coefficients in (\ref{eq:flexi})
can be again implemented as a smooth function in $A$,
\begin{equation}
\label{eq:flexi-adep}
\xi =\lambda_{\xi}+\gamma_{\xi}\,A^{\delta_{\xi}}\,,
\end{equation}
where $\xi=\{n_{q,g},\alpha_{q,g},\beta_{q,g}, n'_{q,g}, \epsilon_{q,g}\}$, and the
$\lambda_{\xi}$, $\gamma_{\xi}$, and $\delta_{\xi}$ need to be determined by the fit.
Allowing for a completely unconstrained $A$ dependence in (\ref{eq:flexi-adep})
would require too many parameters, which in turn would be only 
poorly constrained by data.
Some further guidance is provided by the requirement that nuclear effects should 
vanish as $A\rightarrow 1$, and the nFFs should turn into the well-known vacuum FFs.
Since it is not entirely clear yet that this limit is smooth for small $A\lesssim 4$,
we do not impose the additional constraint that $W_{q,g}^H$ approaches
a delta function $\delta (1-y)$ as $A\rightarrow 1$.
Nevertheless, in most cases $\lambda_{\xi}$ can be set to zero or unity
without damaging the quality of the global fit,
and the corresponding values of $\gamma_{\xi}$ stay very close to zero
as required by a smooth limit if $A\rightarrow 1$.
In addition, some coefficients show a clear preference for a nuclear size 
dependence close to $A^{2/3}$. The assumption of this simple 
geometrical behavior with one common exponent $\delta_{\xi}$ for all the coefficients
in (\ref{eq:flexi-adep}) does not change the quality of the fit.
In total this leaves 14 free parameters to be determined by the fit. 

%%%%%%%%%%%%%%%%%%%%%%%%%%%%%%%%%%%%%%%%%%%%%%%%%%%%
\subsection{\label{sec:pion} NLO analysis of  pion nFFs}
%%%%%%%%%%%%%%%%%%%%%%%%%%%%%%%%%%%%%%%%%%%%%%%%%%%%
%
The free parameters in Eqs.~(\ref{eq:flexi}) and (\ref{eq:flexi-adep}),
as well as for the simplified fit presented in the previous Subsection,
are determined by a standard $\chi^2$ minimization for $N$ data points, where
\begin{equation}
\label{eq:chi2}
\chi^2=\sum_{i=1}^N \frac{(T_i-E_i)^2}{\delta E_i^2}\,.
\end{equation}
$E_i$ is the measured value of a given observable,
$\delta E_i$ the error associated with this measurement, and
$T_i$ denotes the corresponding theoretical estimate for a
given set of parameters.
As often, we take statistical and systematical errors in quadrature 
in $\delta E_i$. Since this is the first attempt of an extraction 
of nFFs from a global QCD analysis, we refrain from a more sophisticated 
treatment of experimental uncertainties. In any case, full information 
on error correlation matrices is not available for most of the data sets 
being analyzed here.
%
%%%%%%%%%%%%%%
% TABLE 2
%%%%%%%%%%%%%%
%
\begin{table*}[th]
\caption{\label{tab:flexi-par}Coefficients parametrizing the 
weight functions $W_{q,g}^{\pi}$ in Eq.~(\ref{eq:flexi})
at the input scale $Q_0=1\,\mathrm{GeV}$ for different nuclei $A$.
The 14 fitted parameters $\lambda_{\xi}$, $\gamma_{\xi}$, and $\delta_{\xi}$ 
in Eq.~(\ref{eq:flexi-adep}) are given in the bottom half of the table.}
\begin{ruledtabular}
\begin{tabular}{cccccccccc}
 A  & $n'_q$ &$\epsilon_q=\epsilon_g$& $n_q$  &$\alpha_q$&$\beta_q$& $n'_g$ 
& $n_g$  &$\alpha_g$&$\beta_g$\\ \hline
 He &  0.949 &   0.002  &  0.024 &  24.18  & 26.80  &  1.042 & -0.131 &  19.50  & 46.24  \\ 
 Ne &  0.863 &   0.006  &  0.065 &  23.54  & 26.13  &  1.114 & -0.351 &  19.21  & 41.78  \\ 
 Kr &  0.668 &   0.015  &  0.157 &  22.10  & 24.62  &  1.276 & -0.849 &  18.56  & 31.72  \\ 
 Xe &  0.564 &   0.020  &  0.206 &  21.33  & 23.81  &  1.362 & -1.115 &  18.21  & 26.33  \\ 
 Au &  0.439 &   0.026  &  0.265 &  20.41  & 22.85  &  1.466 & -1.433 &  17.79  & 19.90  \\ \hline
$\lambda_{\xi}$ & 1      & 0     & 0     &24.56&27.20& 1     & 0      &19.67 & 48.88 \\ 
$\gamma_{\xi}$  &-0.022 & 0.001& 0.010&-0.161&-0.169& 0.018&-0.056 &-0.073 &-1.126\\  
$\delta_{\xi}$  & 0.615 & 0.615& 0.615& 0.615& 0.615& 0.615& 0.615 & 0.615 &0.615\\
\end{tabular}
\end{ruledtabular}
\end{table*}

%%%%%%%%%%%%%%
% TABLE 3
%%%%%%%%%%%%%%
%
\begin{table}[th]
\caption{\label{tab:flexi-chi2}Data sets included in the NLO global
analysis of pion nFFs, the individual $\chi^2$ values for each set,
and the total $\chi^2$ of the fit.}
\begin{ruledtabular}
\begin{tabular}{lccccr}
                               &   &   & Data   & Data   &          \\ 
Experiment                     & A & H & type   & points & $\chi^2$ \\ \hline
HERMES \cite{Airapetian:2007vu}& He,Ne,Kr,Xe&$\pi^+$ &$z$&36     &  39.3        \\
                               & &  $\pi^-$ &$z$   &     36      &  23.0        \\
                               & &  $\pi^0$ &$z$   &     36      &  27.4        \\
                               & &  $\pi^+$ &$x$   &     36      &  69.4        \\
                               & &  $\pi^-$ &$x$   &     36      &  55.4        \\
                               & &  $\pi^0$ &$x$   &     36      &  49.7        \\    
                               & &  $\pi^+$ &$Q^2$ &     32      &  21.0        \\
                               & &  $\pi^-$ &$Q^2$ &     32      &  27.1        \\
                               & &  $\pi^0$ &$Q^2$ &     32      &  34.7        \\
PHENIX \cite{ref:phenix}           & Au &$\pi^0$&$p_T$ &     22      &  13.7        \\
STAR (prel.) \cite{ref:star-thesis}   & Au &$\pi^0$&$p_T$ & 13  &  12.8        \\
STAR \cite{ref:star}             & Au &$\pi^{\pm}$ &$p_T$&  34       &  22.5         \\ \hline
Total                         &    &           &     &  381     &  396.0 \\
\end{tabular}
\end{ruledtabular}
\end{table}
The results of a fit based on the ansatz for the weight functions
$W_{q,g}^\pi$ presented in Eqs.~(\ref{eq:flexi}) and (\ref{eq:flexi-adep}) 
is summarized in Tabs.~\ref{tab:flexi-par} and \ref{tab:flexi-chi2}.
In Table \ref{tab:flexi-par} 
we collect the values of the coefficients parametrizing the 
weight functions $W_{q,g}^{\pi}$ in Eq.~(\ref{eq:flexi}) for
different nuclei and the fitted parameters 
$\lambda_{\xi}$, $\gamma_{\xi}$, and $\delta_{\xi}$ in Eq.~(\ref{eq:flexi-adep}).
Table~\ref{tab:flexi-chi2} shows the partial contributions to $\chi^2$
for each set of data included in the fit.
We obtain an overall $\chi^2=396.0$ for 381 data points included in the 
analysis with 14 free parameters, resulting in an excellent $\chi^2/d.o.f=1.08$.
Recall that the parameters of the simplified, three parameter fit
based on Eqs.~(\ref{eq:toy}) and (\ref{eq:toy-adep}), 
which yields $\chi^2/d.o.f.\simeq 2$, were already presented
in Tab.~\ref{tab:toypara}.

The obtained smooth $A$ dependence of the coefficients parametrizing the 
weight functions $W_{q,g}^{\pi}$ in Eq.~(\ref{eq:flexi}) is 
illustrated in Fig.~\ref{fig:paraplot}. Although not explicitly enforced
in Eq.~(\ref{eq:flexi-adep}), $n_{q,g}'$ approaches unity and
both $n_{q,g}$ and $\epsilon_{q,g}'$ tend to zero as $A\rightarrow 1$,
as required by the vanishing of nuclear effects in that limit.
As expected from the qualitative discussion of the data in Sec.~\ref{sec:data},
the pattern of medium induced modification is rather different for 
quarks and for gluons. Most importantly, the normalizations 
$n'_{q}$ and $n'_{g}$ of the Dirac delta term in the convolution weights 
(\ref{eq:flexi}) have opposite signs, leading to suppression for the 
quark and enhancement for the gluon nFFs with respect to the 
vacuum FFs. The first term on the right-hand-side of Eq.~(\ref{eq:flexi})
also shows an opposite trend for quarks and gluons but influences mainly the
small $z$ behavior of the resulting $D^{\pi}_{q/A}$ and $D^{\pi}_{g/A}$
which will be presented at the end of this Subsection.
%
%%%%%%%%%%%%%%%%%%%%%%%%%%%%%%%
% FIGURE 1: A-DEP OF PARAMETERS
%%%%%%%%%%%%%%%%%%%%%%%%%%%%%%%
%
\begin{figure*}[ht]
\begin{center}
\vspace*{-0.6cm}
\epsfig{figure=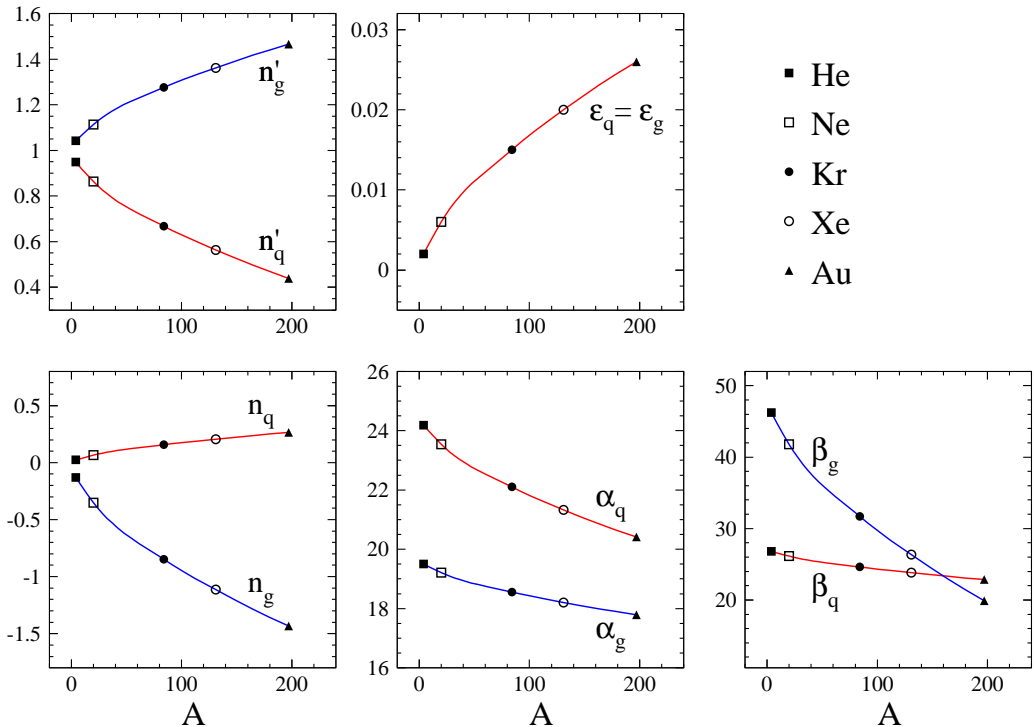,width=0.85\textwidth}
\end{center}
\vspace*{-0.5cm}
\caption{\label{fig:paraplot}$A$ dependence of the coefficients parametrizing the 
weight functions $W_{q,g}^{\pi}$ in Eqs.~(\ref{eq:flexi}) and (\ref{eq:flexi-adep}).}
\end{figure*}
%

%%%%%%%%%%%%%%%%%%
% FIGURE 2: SIDIS
%%%%%%%%%%%%%%%%%%
\begin{figure*}[th]
\begin{center}
\vspace*{-0.6cm}
\epsfig{figure=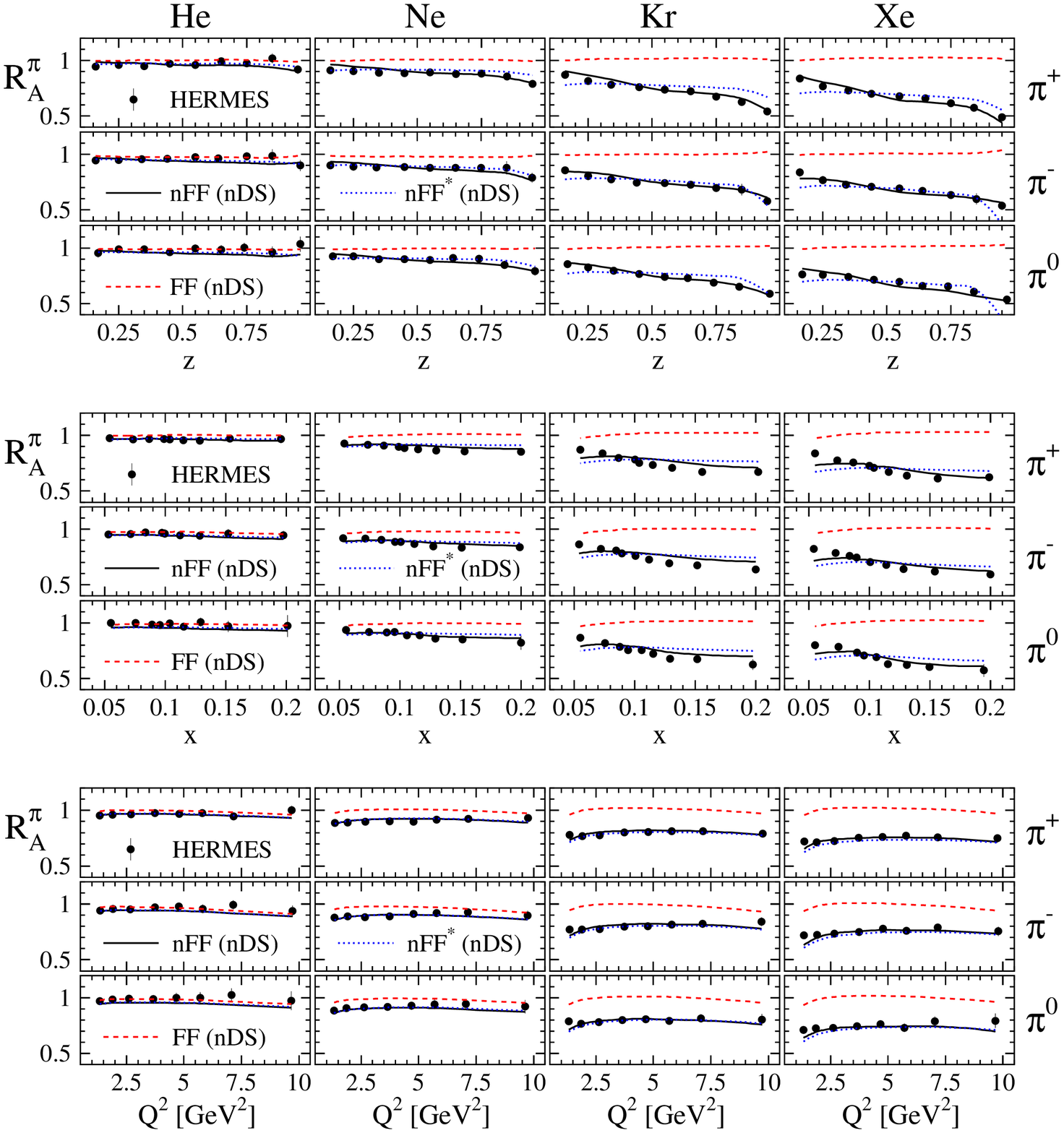,width=0.875\textwidth}
\end{center}
\vspace*{-1.9cm}
\caption{$R_A^{\pi}$ in SIDIS for different nuclei in bins of $z$ (upper panel),
$x$ (middle panel), and $Q^2$ (bottom panel) as measured by HERMES 
\cite{Airapetian:2007vu}. 
The solid lines correspond to the results of our optimum fit for nFFs using
the nDS medium modified parton densities \cite{deFlorian:2003qf}. 
The corresponding fit based on the simple nFF* ansatz in Eq.~(\ref{eq:toy})
is shown as dotted lines.
The dashed lines are estimates assuming the nDS medium modified PDFs 
but standard DSS vacuum FFs \cite{deFlorian:2007}.
\label{fig:sidis}}
\end{figure*}
In Fig.~\ref{fig:sidis}, the measured multiplicity ratios $R_A^{\pi}$ 
for charged and neutral pions \cite{Airapetian:2007vu} are shown
as a function of $z$, $x$, and $Q^2$ for different target nuclei $A$. 
To demonstrate the significance of medium modifications in the hadronization
process, the dashed lines correspond to calculations of $R_A^{\pi}$ 
at NLO accuracy using nPDFs from nDS \cite{deFlorian:2003qf} but
standard DSS fragmentation functions \cite{deFlorian:2007}.
All computations of multiplicities on a deuterium target in the
denominator of Eq.~(\ref{eq:multiplicity}) are performed with
MRST PDFs \cite{ref:unpolpdf} and DSS \cite{deFlorian:2007} FFs.
As anticipated in Sec.~\ref{sec:data}, initial-state nuclear effects 
approximately cancel in $R_A^{\pi}$, and the results computed in this way
are very close to unity for all kinematic distributions, in sharp contrast
to data. Very similar results are obtained if other current sets of 
nPDFs \cite{Hirai:2007sx,Eskola:2009uj} are used.

Notice that despite cancellations of the initial-state effects associated to the 
nPDFs, the measured medium induced modifications of the pion multiplicities
can be as large as a fifty percent effect for the heavier nuclei.
They show a non-trivial $z$ and $x$ dependence, increase 
with the nuclear mass number $A$, and are most conspicuous for larger 
hadron energy fractions $z$ and larger $x$, i.e., smaller $\nu$. 
The dependence on $Q^2$, displayed in the bottom panel of
Fig.~\ref{fig:sidis}, is comparatively flat but noticeable.
Several models proposed to estimate medium induced effects in the hadronization process
reproduce some of the features of the data. However, the full kinematical 
dependence of $R^H_A(\nu,Q^2,z,p_T^2)$ is still a challenging issue; 
see Refs.~\cite{Arleo:2008dn,Accardi:2009qv} and references therein.

%%%%%%%%%%%%%%%%%%
% FIGURE 3: PHENIX
%%%%%%%%%%%%%%%%%%
\begin{figure}[th]
\begin{center}
\vspace*{-0.6cm}
\epsfig{figure=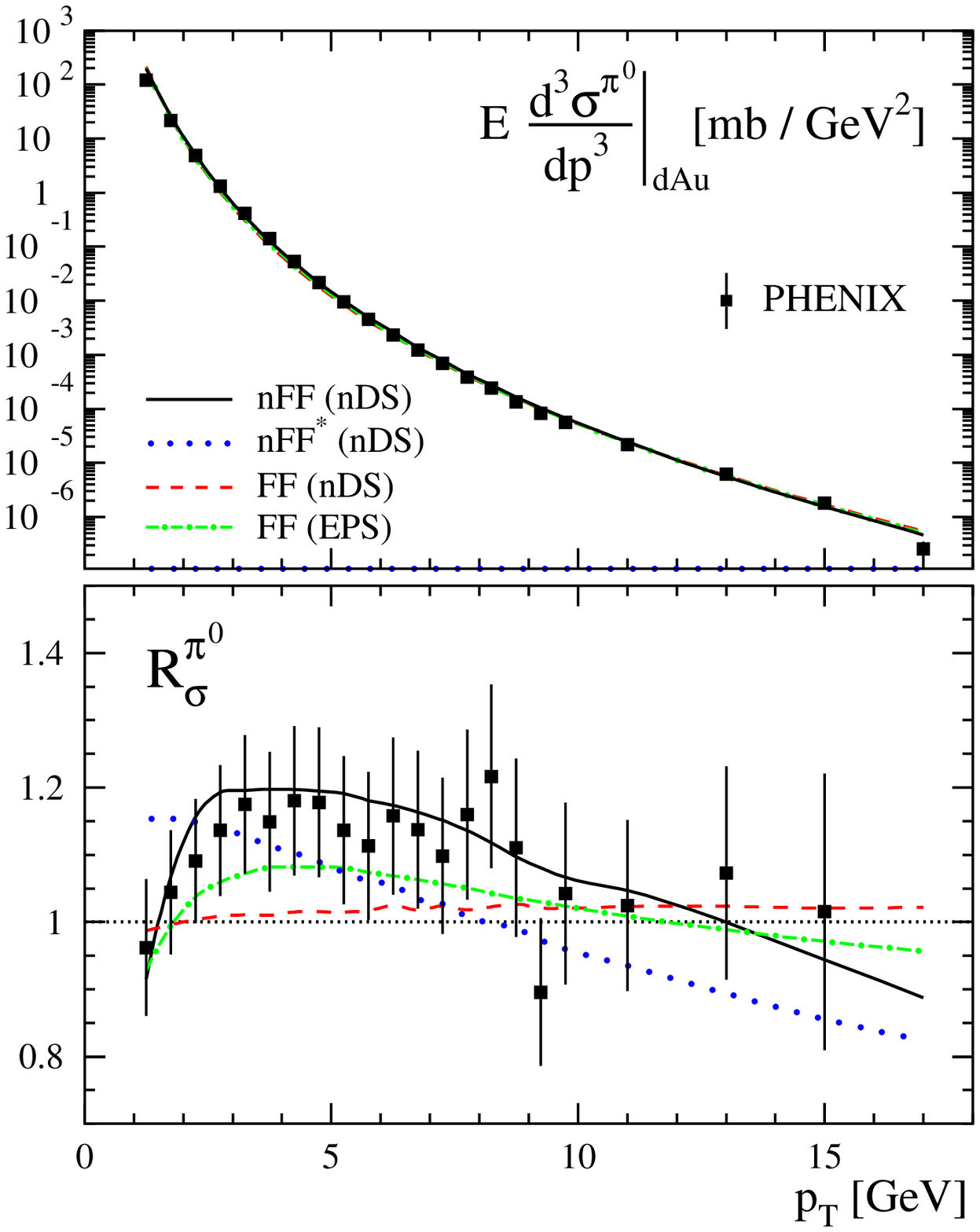,width=0.5\textwidth}
\end{center}
\vspace*{-0.7cm}
\caption{Upper panel: comparison of the PHENIX data for neutral pion production 
in $dAu$ collisions at mid-rapidity \cite{ref:phenix} with NLO estimates obtained with various combinations
of nPDFs, FFs, and nFFs. 
The solid and dotted line correspond to the results of our optimum and simple
three parameter fit for nFFs, respectively, using the nDS nPDFs \cite{deFlorian:2003qf}.
Results based on standard DSS FFs \cite{deFlorian:2007} are shown as dashed and dot-dashed lines
for nDS \cite{deFlorian:2003qf} and EPS \cite{Eskola:2009uj} nPDFs, respectively.
Lower panel: same as above but now for the ratio $R^{\pi}_{\sigma}$ 
defined in Eq.~(\ref{eq:rhicratio}).
\label{fig:phenix}}
\end{figure}
The results of the global analysis of nFFs based on the convolution approach
are shown as dotted and solid lines for the naive three parameter
and the refined ansatz for the weight functions $W^{\pi}_{q,g}$
introduced in Sec.~\ref{sec:weight}, respectively.
Both fits give a much superior description of the full kinematic dependence of the
HERMES data than an approach which ignores final-state nuclear effects.
Even the simple ansatz is doing surprisingly well in reproducing the general trend
of the data, with tensions mainly for the $x$ and $z$ differential yields and 
larger nuclei. 
It is interesting to notice that there seems to be no visible conflict between the standard 
$Q^2$ dependence assumed for the nFFs in our fit and the data.
In this respect, there have been many interesting suggestions and model dependent
calculations at the LO level, motivating the use of medium modified evolution equations,
see \cite{Armesto:2007dt} and references therein. However, in the range of $Q^2$ covered
by present SIDIS data, there is no evidence for a significant departure from standard
timelike evolution equations and kernels in Eq.~(\ref{eq:singevol}) and (\ref{eq:pmatrix}),
respectively.

Even though the full NLO framework for the $p_T$ dependent hadron yields in
SIDIS processes is available \cite{Daleo:2004pn}, and
very precise experimental studies have been presented recently \cite{Airapetian:2007vu}, 
we choose to work only with $p_T$ integrated $R^H_A$ for the time being.
The dependence of the data on this variable is rather weak, and the $p_T$ values 
accessible so far are at the limit of a perturbative treatment. 

%%%%%%%%%%%%%%%%%%
% FIGURE 4: STAR0
%%%%%%%%%%%%%%%%%%
\begin{figure}[th]
\begin{center}
\vspace*{-0.6cm}
\epsfig{figure=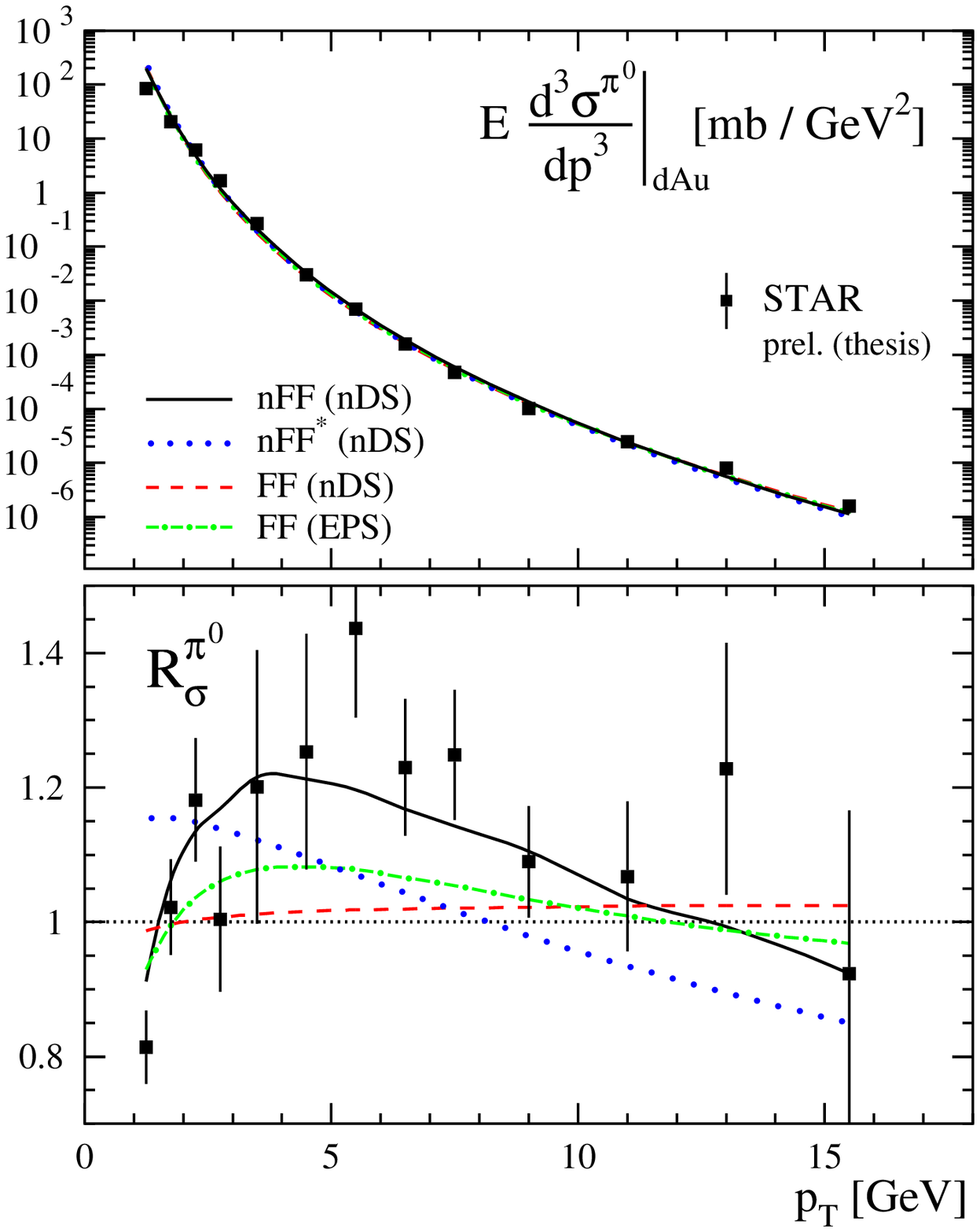,width=0.5\textwidth}
\end{center}
\vspace*{-0.7cm}
\caption{The same as in Fig. \ref{fig:phenix} but now for neutral pion data obtained by STAR
\cite{ref:star-thesis}.
\label{fig:star0}}
\end{figure}
In Figs.~\ref{fig:phenix}, \ref{fig:star0}, and \ref{fig:starpm} we show both the invariant 
$dAu$ cross sections for neutral and charged pion production at RHIC as a function of $p_T$, 
and, in the lower panels, the ratios $R^h_{\sigma}$ to the $pp$ yields, 
defined in Eq.~(\ref{eq:rhicratio}). 
Due to the steep fall of the cross sections with $p_T$ over 
several orders of magnitude in the range shown in Figs.~\ref{fig:phenix}-\ref{fig:starpm},
it is hard to see any differences between the data 
and the NLO calculations based on standard or medium modified FFs.
They become clearly visible, however, in terms of the ratios $R^h_{\sigma}$
and can be as large as about 20\%.

%%%%%%%%%%%%%%%%%%%%%%%
% FIGURE 5: STAR PI+/-
%%%%%%%%%%%%%%%%%%%%%%%
\begin{figure*}[t]
\begin{center}
\vspace*{-0.6cm}
\epsfig{figure=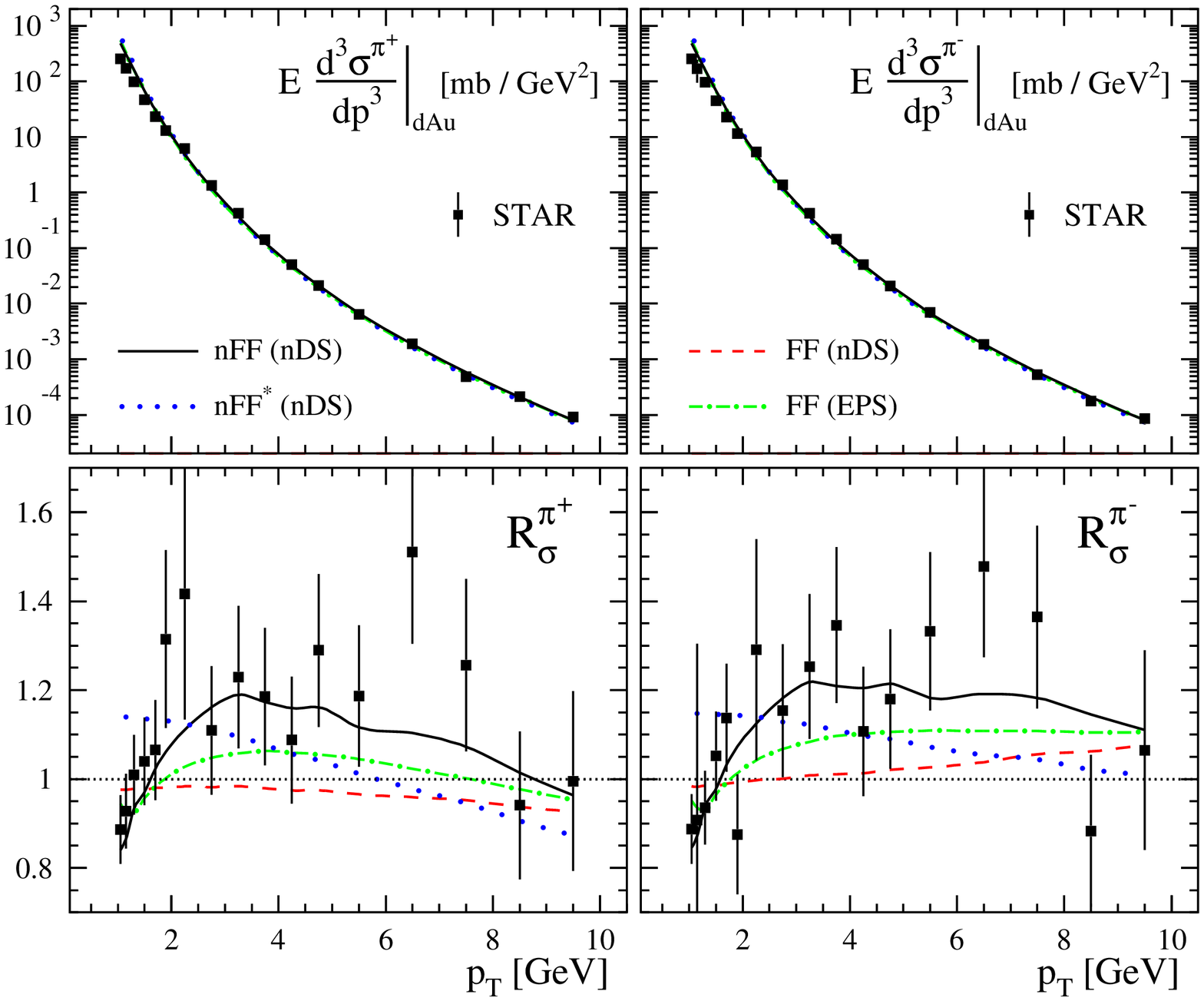,width=0.75\textwidth}
\end{center}
\vspace*{-0.7cm}
\caption{The same as in Figs.~\ref{fig:phenix} and \ref{fig:star0}
but now for charged pion data from STAR \cite{ref:star}.
\label{fig:starpm}}
%\end{figure*}
%%%%%%%%%%%%%%%%%%%
% FIGURE6: <z>, etc
%%%%%%%%%%%%%%%%%%%
%\begin{figure*}[t]
\begin{center}
\vspace*{-0.6cm}
\epsfig{figure=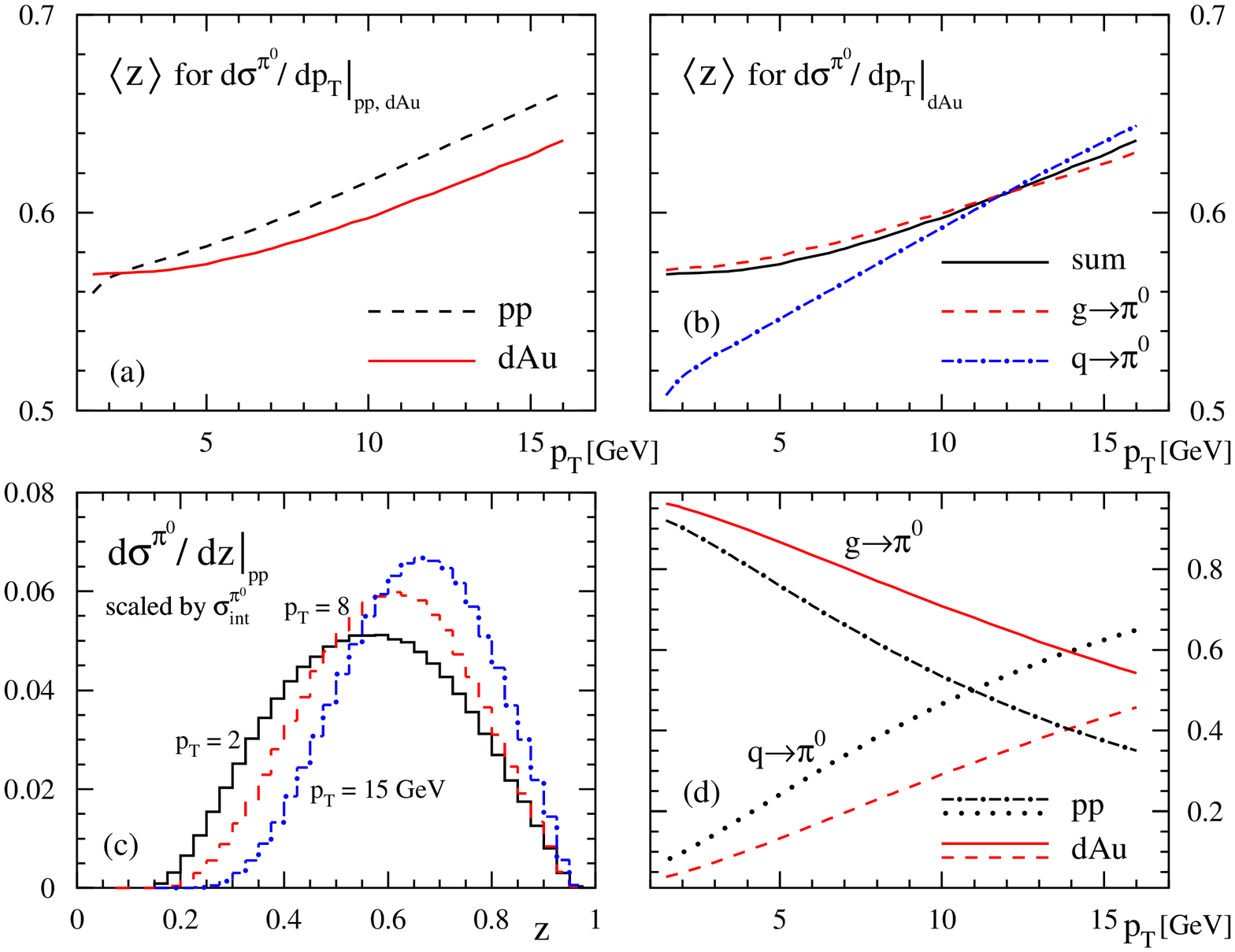,width=0.83\textwidth}
\end{center}
\vspace*{-0.7cm}
\caption{(a) The mean value of $z$ probed in $pp$ and $dAu$ collisions as
a function of $p_T$. (b) The mean value of $z$ in $dAu$ collisions
for both quark and gluon contributions. (c) Histograms of the $z$ distribution 
for three representative $p_T$ values. (d) The relative contributions of quark 
and gluon fragmentation processes to the $\pi^0$ production cross section in
$pp$ and $dAu$ collisions.
\label{fig:meanz}}
\end{figure*}
At variance with $R_A^{\pi}$ in SIDIS, where initial-state medium effects, and hence the differences
associated to the choice of nPDFs, cancel to a large extent, they may remain significant for
$dAu$ data. Thus, estimates of the medium induced modifications for fragmentation functions
could in principle depend on the choice of nPDFs.
To estimate their impact on the $dAu$ cross sections and
on the ratios $R^h_{\sigma}$, we use two different sets of nPDFs, 
nDS \cite{deFlorian:2003qf} and EPS \cite{Eskola:2009uj}, 
along with the well-known standard FFs of DSS for our calculations  
shown in Figs.~\ref{fig:phenix}-\ref{fig:starpm}.
As can be seen, both results do not differ too much and fall short of describing the
data satisfactorily, leaving room for improvement due to medium
induced effects in the hadronization process. 
The EPS set reproduces the trend of the data better than nDS, in particular for
neutral pions shown in Figs.~\ref{fig:phenix} and \ref{fig:star0},
which extend to somewhat larger values of $p_T$ than the charged pion data 
in Fig.~\ref{fig:starpm}. Notice, however, that the $dAu$ data were
included in the EPS analysis of nPDFs, but assuming nuclear effects in the hadronization process 
to be negligible \cite{Eskola:2009uj}.
Although the modification introduced in the EPS nPDFs helps to describe the $dAu$ data
better \cite{Eskola:2009uj}, neglecting final-state nuclear effects is clearly not advisable in view
of their significant impact on $R_A^{\pi}$ in SIDIS demonstrated in Fig.~\ref{fig:sidis}.

The results of our global fit of nFFs are shown in Figs.~\ref{fig:phenix}-\ref{fig:starpm}
as solid and dotted lines, corresponding to our optimum and simplified
ansatz for the weight function $W_{q,g}^{\pi}$
introduced in Sec.~\ref{sec:weight}, respectively.
Here, the naive three parameter ansatz for $W_{q,g}^{\pi}$ fails to reproduce the $p_T$ dependence of 
the $dAu$ data, and the greater flexibility of Eqs.~(\ref{eq:flexi}) and (\ref{eq:flexi-adep})
is clearly needed and leads to a significant improvement of the fit. 
The simultaneous description of SIDIS and $dAu$ data requires to have the correct balance between
quark and gluon contributions in the fragmentation process, which is strongly $p_T$ and,
hence, $z$ dependent.

An important difference between SIDIS and $dAu$ data is that 
in the latter case the cross sections sample contributions from a wide range in $z$.
Consequently, the deconvolution of the medium induced effects is less transparent.
In order to provide a better insight into the sensitivity 
of the RHIC measurements to the fragmentation process, 
we show in Fig.~\ref{fig:meanz}~(a) the mean value of the
hadron's fractional momentum $z$ 
probed in $pp$ and $dAu$ collisions as a function of $p_T$.
There are several ways to estimate an average $\langle z\rangle$.
We define it in the standard way by evaluating the convolutions
in the factorized expression for the $p_T$ dependent cross section \cite{ref:pionnlo}
with an extra factor of $z$ in the integrand, divided by the
cross section itself \cite{ref:guzey}, i.e., schematically we use
\begin{equation}
\label{eq:meanz}
\langle z\rangle \equiv \frac{ \int dz\, z\,\frac{d\sigma^H}{dz dp_T}}
{\int dz  \frac{d\sigma^H}{dz dp_T}}\,.
\end{equation}
Here, $d\sigma^H/dzdp_T$ contains the appropriate convolutions of the parton densities
and fragmentation functions with the partonic hard scattering cross sections.

Panel (b) of Fig.~\ref{fig:meanz} shows the individual $\langle z\rangle$ 
for quark and gluon fragmentation processes, simply referring
to the contributions in the $dAu$ cross section proportional
to either $D_{q/Au}^{\pi}$ or $D_{g/Au}^{\pi}$. 
Beyond the LO, this separation involves some arbitrariness and 
depends on the choice of the factorization scheme.
In addition, primary partons created in the hard scattering
may radiate off secondary partons of a different 
species which in turn fragment into the observed hadron.
Figure~\ref{fig:meanz}~(c) shows histograms of the $z$ distribution 
for three representative values of $p_T$. In panel (d), we present
the relative contributions of quark 
and gluon fragmentation processes to the $\pi^0$ production cross section in
$pp$ and $dAu$ collisions.

As can be seen in Fig.~\ref{fig:meanz}, RHIC $pp$ and $dAu$ data are mainly sensitive to 
fairly large values of the momentum fraction taken by the hadron $H$, 
with $\langle z\rangle$ slightly increasing with $p_T$.
However, as panel (c) shows, the cross section samples contributions over a broad
range in $z$, starting at about $z\simeq 0.2$. Notice that below about $p_T=1.5\,\mathrm{GeV}$,
the tail in the $z$ distribution becomes sensitive to values $z\lesssim 0.1$, where
the concept of fragmentation functions breaks down due to finite hadron mass effects and
the singular behavior of the timelike evolution kernels.
It is also worthwhile mentioning that the values of $z$ to which the cross sections are most
sensitive to, i.e., $\langle z\rangle$, depend, of course, on the actual shape 
of the FFs and nFFs assumed in the analysis of $pp$ and $dAu$ collision data, respectively.
Since we anticipate sizable differences between them, the ratios $R^H_{\sigma}$
defined in Eq.~(\ref{eq:rhicratio}) actually sample the nuclear and the vacuum 
fragmentation functions at slightly different values of $z$,
cf.~Fig.~\ref{fig:meanz}~(a). This can have quite some effect on the ratios
$R^H_{\sigma}$ in regions where the fragmentation functions vary rapidly with $z$.

%%%%%%%%%%%%%%%%%%
% FIGURE 7: D_i/A
%%%%%%%%%%%%%%%%%%
\begin{figure*}[t]
\begin{center}
\vspace*{-0.6cm}
\epsfig{figure=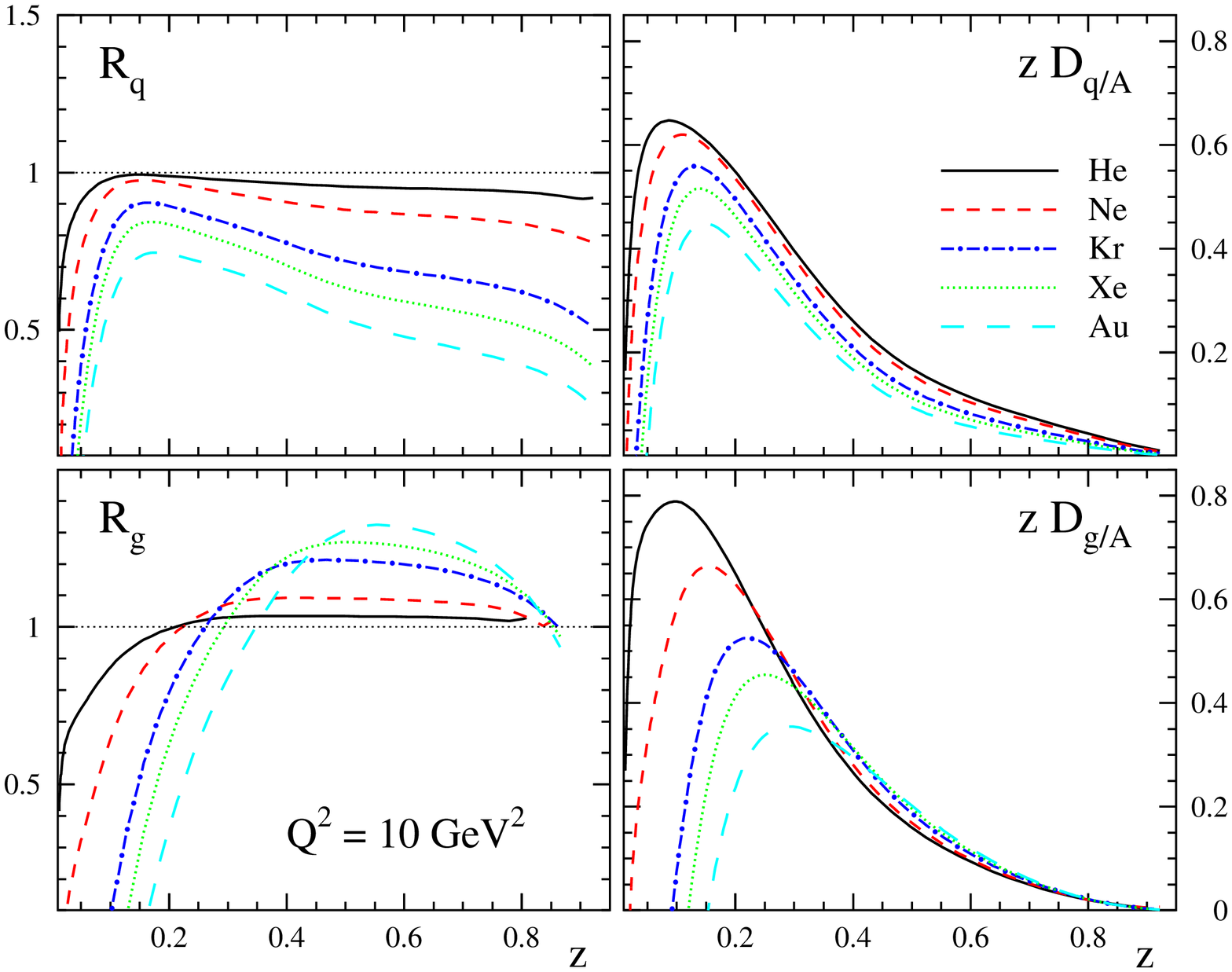,width=0.75\textwidth}
\end{center}
\vspace*{-0.7cm}
\caption{The resulting medium modified NLO fragmentation functions for quarks
and gluons into neutral pions at $Q^2=10\,\mathrm{GeV}^2$ 
for various different nuclei (right panels). 
The left panels show the corresponding
ratios $R_{q,g}^{\pi}$ to the standard DSS FFs as defined in Eq.~(\ref{eq:ff-ratio}).
\label{fig:nffs}}
\end{figure*}
Fig.~\ref{fig:meanz}~(d) demonstrates that for $pp$ collisions gluon fragmentation
is clearly the dominant production mechanism at low values of $p_T$. 
Quark fragmentation contributions increase with $p_T$, and cross the level of 50\%
at $p_T \simeq 10\,\mathrm{GeV}$. Again,
the relative balance between quark and gluon contributions in $dAu$ collisions will 
depend on the extracted medium induced modifications. In our analysis, pion production
is dominantly driven by gluon fragmentation up to the highest values of $p_T$,
about $15\,\mathrm{GeV}$, currently accessible in experiment. As we shall see below,
this is due to the nuclear suppression of $D_{q/A}^H$ on the one hand, and enhancement
of $D_{g/A}^H$ on the other hand.

The resulting modified fragmentation functions and the corresponding ratios 
to the standard DSS FFs,
\begin{equation}
\label{eq:ff-ratio}
R_q^H \equiv \frac{D^H_{q/A}(z,Q^2)}{D^H_q(z,Q^2)},  
\,\,\,R_g^H \equiv \frac{D^H_{g/A}(z,Q^2)}{D^H_g(z,Q^2)},  
\end{equation}
can be found in Fig.~\ref{fig:nffs} at $Q^2=10\,\mathrm{GeV}^2$ for various nuclei.
As expected from the qualitative discussion of the data in Sec.~\ref{sec:data}, 
the pattern of medium induced modifications is rather different for quarks and for gluons.
The dominant role of quark fragmentation in SIDIS leads to a suppression, i.e., $R_q^{\pi}<1$,
increasing with nuclear size $A$ as
dictated by the pattern of hadron attenuation found experimentally, see Fig.~\ref{fig:sidis}.
The enhancement of hadrons observed in $dAu$ collisions for $p_T\lesssim 10\,\mathrm{GeV}$,
see Figs.~\ref{fig:phenix}-\ref{fig:starpm}, along with the dominant role of
gluon fragmentation at low values of $p_T$ explains that $R_g^{\pi}>1$ for $z\gtrsim 0.2$.
Below $z\simeq 0.2$, where all the data we analyze have very little or no constraining
power, both quark and gluon nFFs drop rapidly. For the time being, the behavior in this 
region could easily be an artifact of the currently assumed functional form for the 
weights $W_{q,g}^H$ in Eq.~(\ref{eq:flexi}).

We note that for $z\gtrsim 0.4$ the bulk of the medium induced effects for both quarks
and gluons can be accommodated by the second term in $W_{q,g}^H$ proportional to a
Dirac delta function, which normalization coefficients $n'_q$ and $n'_g$ have to have
opposite signs, reflecting suppression and enhancement, respectively.
The other term in $W_{q,g}^H$ introduces only a small correction in this region, 
but becomes dominant if $z\lesssim 0.4$, especially for $D_{g/A}^H$.

We wish to stress, that even with the limited amount of data available at present,
the successfully performed global analysis of nFFs provides a first non-trivial
indication that the assumed factorization of long- and short-distance physics 
works amazingly well also for SIDIS off nuclei and $dAu$ collisions at RHIC.
The found suppression $R_q^{\pi}<1$, as imposed by SIDIS data, is fully compatible with the complicated pattern of
enhancement and attenuation of hadron yields in $dAu$ collisions at RHIC.
The non-negligible role of quark fragmentation in $dAu$ collisions at 
moderate and large values of $p_T$, along with the larger nuclear effects for
$D_{q/A}^H$ than for $D_{g/A}^H$ at large values of $z$, explains that 
$R_{\sigma}^{\pi}<1$ for $p_T\gtrsim 10\,\mathrm{GeV}$.
In addition, even though SIDIS data are dominantly sensitive to quark fragmentation,
the observed non-trivial $Q^2$ and $x$ dependence of the SIDIS multiplicity ratios
shown in Fig.~\ref{fig:sidis}, depends on the medium induced modifications
of the gluon fragmentation function, which in turn is mainly constrained by $dAu$ data.
The correlation of the $x$ dependence of SIDIS multiplicity rates with the
$z$ dependence of the fragmentation functions is 
induced by the NLO hard scattering coefficient functions which
depend in a non-trivial way on both $x$ and $z$; see, e.g., Ref.~\cite{ref:lambda}.

%%%%%%%%%%%%%%%%%%%%%%%%%%%%%%%%%%%%%%%%%%%%%%%%%%%%%%%%
\subsection{\label{sec:cc} Dependence on different centrality classes}
%%%%%%%%%%%%%%%%%%%%%%%%%%%%%%%%%%%%%%%%%%%%%%%%%%%%%%%%
%
%%%%%%%%%%%%%%%%%%%%%%%%%%%
% FIGURE 8: centrality x-sec
%%%%%%%%%%%%%%%%%%%%%%%%%%%
\begin{figure}[t]
\begin{center}
\vspace*{-0.6cm}
\epsfig{figure=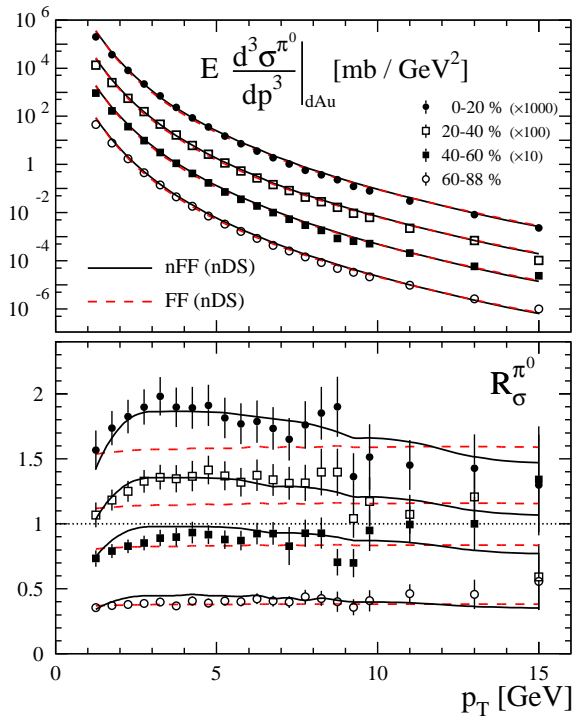,width=0.5\textwidth}
\end{center}
\vspace*{-0.7cm}
\caption{As in Fig.~\ref{fig:phenix} but now separated into different centrality classes
ranging from the most central one, $0-20\%$, to the most peripheral one,
$60-88\%$.
The theoretical estimates obtained for minimum bias events are rescaled by the
ratio ${\cal{C}}$ defined in Eq.~(\ref{eq:centrality}) to account for the different 
centrality classes.
\label{fig:central}}
\end{figure}
All the data on hadron yields in $dAu$ collisions we have discussed and analyzed so far 
correspond to minimum bias events.
Experiments also present their results divided into different centrality classes depending
on how central or how peripheral the collision is in impact parameter space.
In that respect, the nFFs obtained in our global analysis correspond to 
some sort of average of the medium induced effects seen for the 
different degrees of geometrical overlap between the deuteron and the gold nuclei.
One could expect that this average underestimates the medium induced effects 
when the collisions are more central and overestimates them 
for the more peripheral ones. 

To estimate the possible impact of different centrality classes on the
extraction of nFFs, we scale our results obtained for the minimum bias 
cross sections shown in Figs.~\ref{fig:phenix}-\ref{fig:starpm}, by
the average ratio between the measured cross section for 
a given centrality class (c.c.) and the minimum bias (m.b.) sample, i.e.,
\begin{equation}
\label{eq:centrality}
{\cal{C}} \equiv  \left< 
\frac{E\, d^3\sigma^{\pi^0}\!\!/dp^3 \big|^{c.c.}_{dAu}}
{E\, d^3\sigma^{\pi^0}\!\!/dp^3\big|^{m.b.}_{dAu}}  \right>_{c.c.}.
\end{equation}
The ratio (\ref{eq:centrality}) is a simple estimate of the fraction of 
events selected by a given centrality cut.

%%%%%%%%%%%%%%%%%%%%%%%%%%
% FIGURE 9: centrality II
%%%%%%%%%%%%%%%%%%%%%%%%%%
\begin{figure}[t]
\begin{center}
\vspace*{-0.6cm}
\epsfig{figure=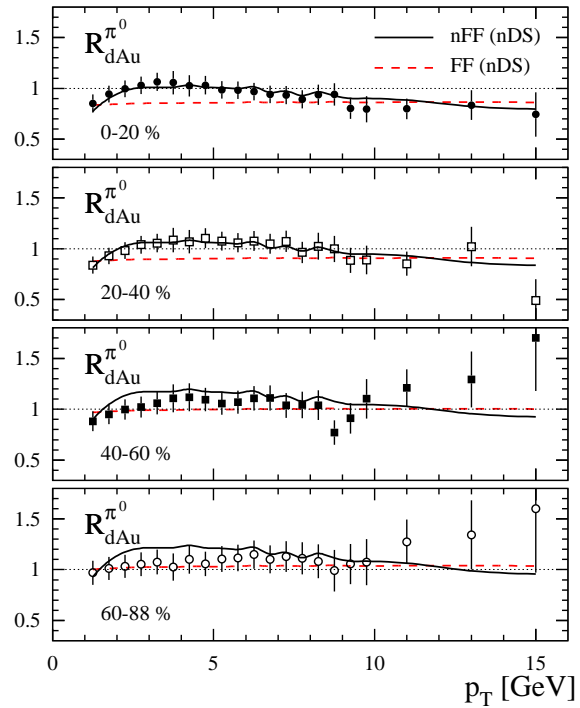,width=0.5\textwidth}
\end{center}
\vspace*{-0.7cm}
\caption{As in the lower panel of Fig.~\ref{fig:central} but now in terms of
the experimentally determined nuclear modification factors $R_{dAu}^{\pi^{0}}$
\cite{ref:phenix}.
\label{fig:rcentral}}
\end{figure}
As a representative example, we show in Fig.~\ref{fig:central} 
the comparison between the PHENIX data for four different centrality classes \cite{ref:phenix}
and the corresponding minimum bias cross section computed with our nFFs and rescaled
by ${\cal{C}}$. As expected, our nFFs slightly overestimate the data for the 
most peripheral events denoted by $40-60\%$ and $60-88\%$. The agreement with
the more central events is, however, very good, and we do not find the expected
underestimate of nuclear effects. As for the minimum bias cross sections
shown in Figs.~\ref{fig:phenix}-\ref{fig:starpm},
estimates obtained with standard vacuum FFs do not reproduce
the trend of the data very well, except for the most peripheral events.
Results for pion yields in different centrality
classes obtained by the STAR experiment are very similar to those presented in  
Fig.~\ref{fig:central} and hence not shown.
For completeness, Fig.~\ref{fig:rcentral} gives a comparison similar to those 
presented in the lower panel of Fig.~\ref{fig:central} but 
now in terms of the nuclear modification factors. 

%
%%%%%%%%%%%%%%%%%%%%
% TABLE 4: kaon par
%%%%%%%%%%%%%%%%%%%
%
\begin{table*}[thb]
\caption{\label{tab:flexika} Coefficients parametrizing the weight functions $W_{q,g}^K$ in
Eq.~(\ref{eq:flexi}) at the input scale $Q_0 =1\,\mathrm{GeV}$ for different nuclei $A$.
The 13 fitted parameters $\lambda_{\xi}$ and $\gamma_{\xi}$ are given in the bottom
half of the table. Note that the $\epsilon_{\xi}$ are chosen as in Tab.~\ref{tab:flexi-par}.}
\begin{ruledtabular}
\begin{tabular}{cccccccccc}
 A  & $n'_q$ &$\epsilon$& $n_q$  &$\alpha_q$&$\beta_q$& $n'_g$ & $n_g$  &$\alpha_g$&$\beta_g$\\ \hline
 He &   0.957 &  0.003 &   0.024 &  27.95 &  30.16 &  1.038 &  -0.107 &  17.75 &  39.37 \\  
 Ne &   0.884 &  0.007 &   0.065 &  27.37 &  29.55 &  1.103 &  -0.288 &  17.40 &  35.33 \\  
 Kr &   0.719 &  0.017 &   0.156 &  26.08 &  28.17 &  1.248 &  -0.695 &  16.60 &  26.22 \\ 
 Ze &   0.630 &  0.022 &   0.205 &  25.38 &  27.43 &  1.326 &  -0.913 &  16.17 &  21.34 \\   
 Au &   0.525 &  0.028 &   0.264 &  24.55 &  26.55 &  1.419 &  -1.174 &  15.67 &  15.52 \\ \hline  
$\lambda_{\xi}$ & 1      & 0     & 0 &  28.29 &  30.52 & 1      & 0       &  17.96 & 41.76 \\ 
$\gamma_{\xi}$  &-0.0185 & 0.0011& 0.0102&-0.1451&-0.1543& 0.0163&-0.0456 &-0.089 &-1.020\\  
\end{tabular}
\end{ruledtabular}
\end{table*}
%
%%%%%%%%%%%%%%%%%%%%%%%%%%%%%%%%%%%%%%%%%%%%%%%%%%%%%%%%
\subsection{\label{sec:kaons} NLO analysis of kaon nFFs}
%%%%%%%%%%%%%%%%%%%%%%%%%%%%%%%%%%%%%%%%%%%%%%%%%%%%%%%%
%
Even though the available experimental information on kaon production 
in a nuclear environment is much more limited than in the case of pions,
it is interesting to study to what extent the pattern of nuclear modifications
found for pions is similar to those for kaons.
It should be noted that the standard kaon FFs,
which will provide the baseline to analyze medium induced effects,
still suffer from sizable uncertainties \cite{deFlorian:2007}.

In a scenario where the medium induced modifications to the hadronization process are
dominated by partonic mechanisms, and to a first approximation in a convolutional approach,
one expects a very similar pattern of medium modifications for pions and for kaons.
If interactions of the produced hadron or intermediate pre-hadrons with the 
nuclear environment are important, medium induced effects in the production of 
pions and kaons can be significantly different.

Due to the relative scarcity of data with identified kaons, especially in the case of 
$dAu$ collisions where most of the data are taken at $p_T$ values 
below the reach of perturbative QCD methods,  
we cannot proceed with our global analysis as we did pions.
The coefficients of a completely flexible parametrization for the weights $W_{q,g}^K$ 
like in Eq.~(\ref{eq:flexi}) would not be well constrained by the data.
Therefore, our strategy is as follows: we start by imposing a more constrained
ansatz for the weights $W_{q,g}^K$ where we assume some of the features
of the medium induced effects found for the pion nFFs.
More specifically, we keep the same functional form for the weights as in 
Eq.~(\ref{eq:flexi}), with the same nuclear $A$ dependence as in Eq.~(\ref{eq:flexi-adep}).
We also take over the simplifying assumption $\epsilon_q=\epsilon_g$, which we have chosen
for pion nFFs in Secs.~\ref{sec:weight} and \ref{sec:pion} 
after checking that more flexible options did not lead to a significant
improvement of the quality of the fit.
In addition, we try to set the exponents $\delta_{\xi}$ governing the
$A$ dependence in Eq.~(\ref{eq:flexi-adep}) to the values preferred by the pion data.
%
%%%%%%%%%%%%%%%%%%%%%
% TABLE 4: kaon chi2
%%%%%%%%%%%%%%%%%%%%%
%
\begin{table}[t]
\caption{\label{tab:chi2-kaon} Data sets included in the NLO global
analysis of kaon nFFs, the individual $\chi^2$ values for each set,
and the total $\chi^2$ of the fit.}
\begin{ruledtabular}
\begin{tabular}{lccccr}
                               &   &   & Data   & Data   &          \\ 
Experiment                     & A & H & type   & points & $\chi^2$ \\ \hline
HERMES \cite{Airapetian:2007vu}& He,Ne,Kr,Xe&  $K^+$ &$z$   &     36      &    61.3     \\
                                & &  $K^-$ &$z$   &     36      & 91.7\\
                                & &  $K^+$ &$x$   &     36      &  67.2      \\
                                & &  $K^-$ &$x$   &     36      &  128.5        \\
                                & &  $K^+$ &$Q^2$ &     32      &  38.4        \\
                                & &  $K^-$ &$Q^2$ &     32      &  50.0     \\
STAR \cite{ref:star}       & Au &$K^+$&$p_T$ &     5        &   5.4      \\
                               & Au &$K^-$&$p_T$ &     5        &  4.6     \\ \hline
Total                          &    &           &     &  218   &  447.1 \\
\end{tabular}
\end{ruledtabular}
\end{table}

Next, we can assess how much variation of the other parameters relative to the 
results for the pion nFFs listed in Tab.~\ref{tab:flexi-par}
is required to reproduce the main features of the kaon data. 
We find that allowing for an up to 20\% variation 
yields a more than reasonable agreement with all available data sets,
with most of the parameters staying within a 10\% variation
of their counterparts describing the pion nFFs.
In Tab.~\ref{tab:flexika} we list the values 
of the coefficients parametrizing the weight functions $W^K_{q,g}$ 
in Eq.~(\ref{eq:flexi}) for different nuclei and the 13 fitted parameters
$\lambda_{\xi}$ and $\gamma_{\xi}$ in Eq.~(\ref{eq:flexi-adep});
the $\delta_{\xi}$ are fixed to the values given in Tab.~\ref{tab:flexi-par}.

The overall quality of the fit is summarized in Tab.~\ref{tab:chi2-kaon}. 
The total $\chi^2$ of the fit is 447.1 for 218 data points included in the
global analysis, resulting in $\chi^2/d.o.f.=2.2$. Also given are the
individual contributions to $\chi^2$ for each set of data included in the fit. 
Even though the quality of the fit is not as good as in the case of pions, it reproduces
the data well within their uncertainties and suggests a close similarity 
between the medium induced modifications for pion and kaon fragmentation functions.
As it could be expected, the most significant variations
are found in the parameters related to the gluon fragmentation $D_{g/A}^K$,
which are only poorly constrained by the scarce $dAu$ data.
Note that the largest individual contribution to $\chi^2$ stems from the SIDIS data with
identified $K^-$ where already the vacuum FFs show some tension with the experimental results 
\cite{deFlorian:2007}.

%%%%%%%%%%%%%%%%%%%%%
% FIGURE 10: SIDIS K
%%%%%%%%%%%%%%%%%%%%%
\begin{figure*}[t]
\begin{center}
\vspace*{-0.6cm}
\epsfig{figure=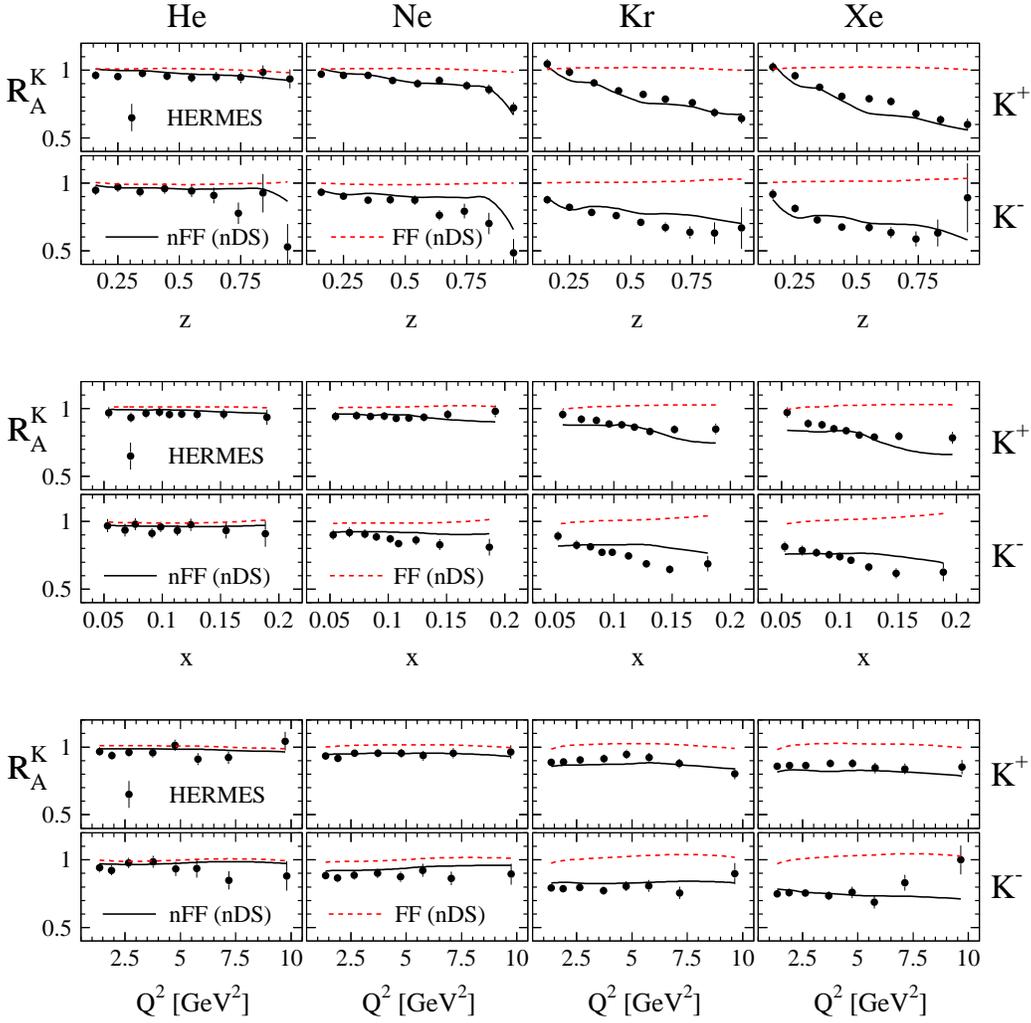,width=0.8\textwidth}
\end{center}
\vspace*{-0.5cm}
\caption{As in Fig.~\ref{fig:sidis} but now for identified $K^{\pm}$.
\label{fig:sidiska}}
\end{figure*}
In Fig.~\ref{fig:sidiska} we compare the result of our fit with the
measured SIDIS multiplicity ratios $R^K_A$ for charged kaons
as a function of $z$, $x$, and $Q^2$ for different target nuclei $A$.
As can be seen, the agreement with the $K^+$ data is reasonably good with
some problems in reproducing the $x$ distributions for heavier nuclei.
Discrepancies are somewhat larger for $K^-$ data as we have already mentioned.
For comparison, we show again the results of a theoretical calculation based
on standard vacuum FFs (dashed lines). As for the pion multiplicity ratios shown in
Fig.~\ref{fig:sidis}, not even the trend of the data can be reproduced
by ignoring medium induced modifications to the hadronization process.
Finally, Fig.~\ref{fig:dauka} shows both the invariant $dAu$ cross section
for charged kaon production as a function of $p_T$, and, in the lower
panels, the ratios $R_{\sigma}^K$ to the corresponding $pp$ yields. One should notice
the limited range in $p_T$ covered by the presently available data which
is certainly at the borderline where pQCD is applicable. Nevertheless,
our fit describes the data well except for the lowest bin in $p_T$.

%%%%%%%%%%%%%%%%%%%%%%%%
% FIGURE 11: kaons STAR
%%%%%%%%%%%%%%%%%%%%%%%%
\begin{figure*}[t]
\begin{center}
\vspace*{-0.6cm}
\epsfig{figure=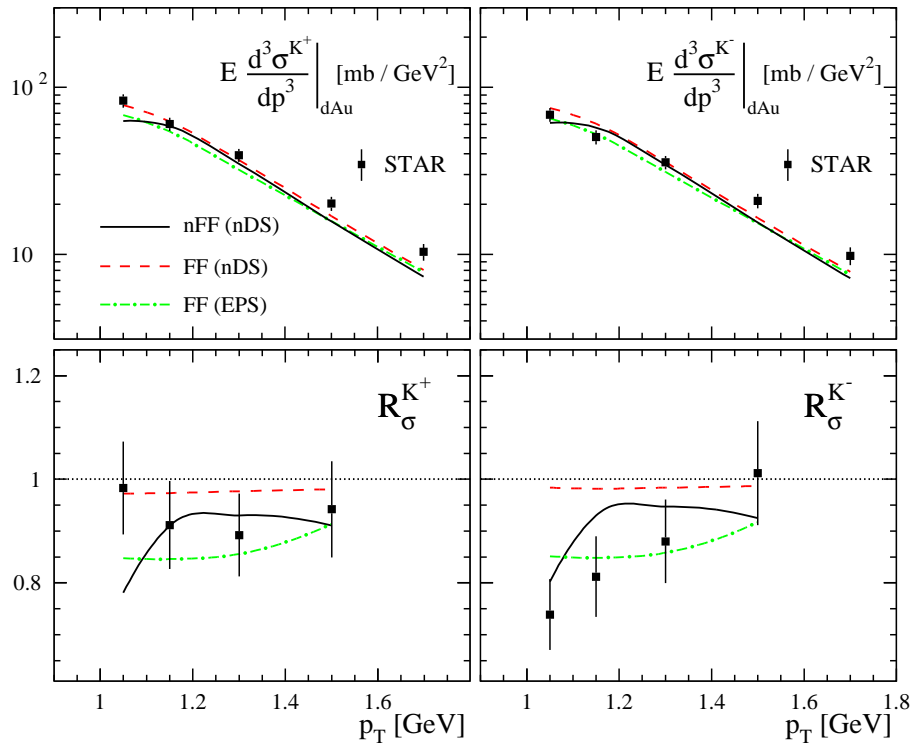,width=0.75\textwidth}
\end{center}
\vspace*{-0.7cm}
\caption{As in Fig.~\ref{fig:starpm} but now for charged kaon production.
\label{fig:dauka}}
\end{figure*}
The availability of both more precise kaon data and more accurate sets of 
vacuum FFs in the future, will help to further investigate the interesting and non-trivial
close relation between the nuclear modifications for pion and kaon fragmentation
functions found in our analysis.

%%%%%%%%%%%%%%%%%%%%%%%%%%%%%%%%%%%%%%%%%%%%%%%%%%%
\section{\label{sec:conclu} Summary and Conclusions}
%%%%%%%%%%%%%%%%%%%%%%%%%%%%%%%%%%%%%%%%%%%%%%%%%%%%
%
We have investigated the possibility of using a fully factorized approach
at NLO accuracy of pQCD, similar to those established for analyses
of standard FFs and PDFs, to describe medium induced effects in the hadronization process. 

To this end, we have explored the concept of modified fragmentation functions 
which effectively accounts for all the distortions in the production of hadrons
caused by the nuclear environment.
These novel distributions, which we denote as nFFs, are completely analogous to 
nuclear PDFs, and, combined with them, allow one to 
treat a large class of hard reactions where a nucleus collides with a lepton, a nucleon, or a very 
light nucleus like a deuteron in a consistent pQCD framework.

We have performed combined NLO fits to the available pion and kaon production data in
SIDIS and in $dAu$ collisions to reveal the non-perturbative
features of the nFFs, exploiting the well established potential of pQCD
to describe hard scattering processes and the wealth of information 
compiled in PDFs and nPDFs over the past years. 
We choose to relate the nFFs to the well-known standard FFs in a convolutional
approach with a great economy of free fit parameters, including a 
smooth dependence on the nuclear size $A$.

The obtained parameterizations of nFFs for pions and kaons \cite{ref:para} rather accurately
reproduce the data and their nuclear $A$ dependence. Our results provide support to the
idea that conventional factorization of short and long-distance physics effects works
to a good approximation also in the nuclear environments studied in our analyses. 
The found pattern of medium induced modification is rather different 
for fragmenting quarks and gluons, where we find suppression and enhancement,
respectively, compared to the vacuum fragmentation functions.
Since our nFFs parametrize the available data
without resorting to a certain model or underlying mechanism for the observed medium
modifications, they may help to further our understanding on hadronization in a nuclear
environment. Predictions based on the observed pattern for quark and gluon nFFs
can be tested by upcoming data from BNL-RHIC and CERN-LHC and perhaps in the future
at an electron-ion collider like the EIC project. This will help to understand
the limits of the factorized approach which is expected to be only an approximation
and may receive corrections beyond the leading twist level or beyond a pQCD approach.

%%%%%%%%%%%%%%%%
\section*{Acknowledgments}
%%%%%%%%%%%%%%%%
%
We warmly acknowledge Elke Aschenauer and Abhay Deshpande for help with 
the HERMES and RHIC data, respectively, and Daniel de Florian for helpful discussions.
This work was partially supported by CONICET, ANPCyT, UBACyT, BMBF, and 
the Helmholtz Foundation.

%%%%%%%%%%%%%%%%%%%%%%%%%%%


\begin{thebibliography}{99}
\bibitem{ref:para} {\sc Fortran} codes of the obtained parameterizations
are available upon request from the authors.
\end{thebibliography}
\end{document}